\documentclass[longauth]{aa}
\usepackage{graphicx}
\usepackage{txfonts}
\usepackage{upgreek}
\usepackage{natbib}
\usepackage{xcolor}
\usepackage[colorlinks=true,linkcolor=blue,citecolor=blue,urlcolor=blue]{hyperref}

\begin{document} 

\title{JOYS: Disentangling the warm and cold material in the high-mass IRAS\,23385+6053 cluster}

\subtitle{}

\author{C. Gieser
	\inst{1,2}
	\and
	H. Beuther\inst{2}
	\and
	E.~F. van Dishoeck\inst{3,1}
	\and
	L. Francis\inst{3}
	\and
	M.~L. van Gelder\inst{3}
	\and
	L. Tychoniec\inst{4}
	\and
	P.~J. Kavanagh\inst{5}
	\and
	G. Perotti\inst{2}
	\and
	A. Caratti o Garatti\inst{6}
	\and
	T.P. Ray\inst{7}
	\and
	P. Klaassen\inst{8}
	\and
	K. Justtanont\inst{9}
	\and
	H. Linnartz\inst{10}
	\and
	W.~R.~M. Rocha\inst{3,10}
	\and
	K. Slavicinska\inst{3,10}
	\and
	L. Colina\inst{11}
	\and
	M. Güdel\inst{12, 2, 13}
	\and
	Th. Henning\inst{2}
	\and
	P.-O. Lagage\inst{14}
	\and
	G. Östlin\inst{15}
	\and
	B. Vandenbussche\inst{16}
	\and
	C. Waelkens\inst{16}
	\and 
	G. Wright\inst{8}
	}

	\institute{Max Planck Institute for Extraterrestrial Physics, Gießenbachstraße 1, 85749 Garching bei M\"unchen, Germany\\
	\email{gieser@mpe.mpg.de}
	\and
	Max Planck Institute for Astronomy, K\"onigstuhl 17, 69117 Heidelberg, Germany
	\and
	Leiden Observatory, Leiden University, PO Box 9513, 2300 RA Leiden, The Netherlands
	\and
	European Southern Observatory, Karl-Schwarzschild-Strasse 2, 85748 Garching bei München, Germany
	\and
	Department of Experimental Physics, Maynooth University-National University of Ireland Maynooth, Maynooth, Co Kildare, Ireland
	\and
	INAF-Osservatorio Astronomico di Capodimonte, Salita Moiariello 16, I-80131 Napoli, Italy
	\and
	Dublin Institute for Advanced Studies, 31 Fitzwilliam Place, D02 XF86 Dublin, Ireland
	\and
	UK Astronomy Technology Centre, Royal Observatory Edinburgh, Blackford Hill, Edinburgh EH9 3HJ, UK
	\and
	Department of Space, Earth and Environment, Chalmers University of Technology, Onsala Space Observatory, 439 92 Onsala, Sweden
	\and
	Laboratory for Astrophysics, Leiden Observatory, Leiden University, PO Box 9513, NL 2300 RA Leiden, The Netherlands
	\and
	Centro de Astrobiologıa (CAB, CSIC-INTA), Carretera de Ajalvir, 8850 Torrejon de Ardoz, Madrid, Spain
	\and
	Department of Astrophysics, University of Vienna, Türkenschanzstr. 17, 1180 Vienna, Austria
	\and
	ETH Zürich, Institute for Particle Physics and Astrophysics, Wolfgang-Pauli-Str. 27, 8093 Zürich, Switzerland
	\and
	Université Paris-Saclay, Université de Paris, CEA, CNRS, AIM, 91191 Gif-sur-Yvette, France
	\and
	Department of Astronomy, Oskar Klein Centre, Stockholm University, 106 91 Stockholm, Sweden
	\and
	Instituut voor Sterrenkunde, KU Leuven, Celestijnenlaan 200D, Bus-2410, 3000 Leuven, Belgium
	}

	\date{Received x; accepted x}

	\abstract
	{High-mass star formation occurs in a clustered mode where fragmentation is observed from an early stage onward. Young protostars can now be studied in great detail with the recently launched \textit{James Webb Space Telescope} (JWST).}
	{We study and compare the warm ($>$100\,K) and cold ($<$100\,K) material toward the high-mass star-forming region IRAS\,23385+6053 (IRAS\,23385 hereafter) combining high angular resolution observations in the mid-infrared (MIR) with the JWST Observations of Young protoStars (JOYS) project and with the NOrthern Extended Millimeter Array (NOEMA) at mm wavelengths at angular resolutions of $\approx$0\farcs2-1\farcs0.}
	{The spatial morphology of atomic and molecular species is investigated by line integrated intensity maps. The temperature and column density of different gas components is estimated using H$_{2}$ transitions (warm and hot component) and a series of CH$_{3}$CN transitions as well as 3\,mm continuum emission (cold component).}
	{Toward the central dense core in IRAS\,23385 the material consists of relatively cold gas and dust ($\approx$50\,K), while multiple outflows create heated and/or shocked H$_{2}$ and show enhanced temperatures ($\approx$400\,K) along the outflow structures. An energetic outflow with enhanced emission knots of $[$Fe\,{\sc ii}$]$ and $[$Ni\,{\sc ii}$]$ hints at $J$-type shocks, while two other outflows have enhanced emission of only H$_{2}$ and $[$S\,{\sc i}$]$ caused by $C$-type shocks. The latter two outflows are also more prominent in molecular line emission at mm wavelengths (e.g., SiO, SO, H$_{2}$CO, and CH$_{3}$OH). Even higher angular resolution data are needed to unambiguously identify the outflow driving sources given the clustered nature of IRAS\,23385. While most of the forbidden fine structure transitions are blueshifted, $[$Ne\,{\sc ii}$]$ and $[$Ne\,{\sc iii}$]$ peak at the source velocity toward the MIR source A/mmA2 suggesting that the emission is originating from closer to the protostar.}
	{The warm and cold gas traced by MIR and mm observations, respectively, are strongly linked in IRAS\,23385. The outflows traced by MIR H$_{2}$ lines have molecular counterparts in the mm regime. Despite the presence of multiple powerful outflows that cause dense and hot shocks, a cold dense envelope still allows star formation to further proceed. To study and fully understand the spatially resolved MIR-properties, a representative sample of low- and high-mass protostars has to be probed by JWST.}


	\maketitle
	
\section{Introduction}\label{sec:intro}

\begin{figure*}[!htb]
\centering
\includegraphics[width=0.99\textwidth]{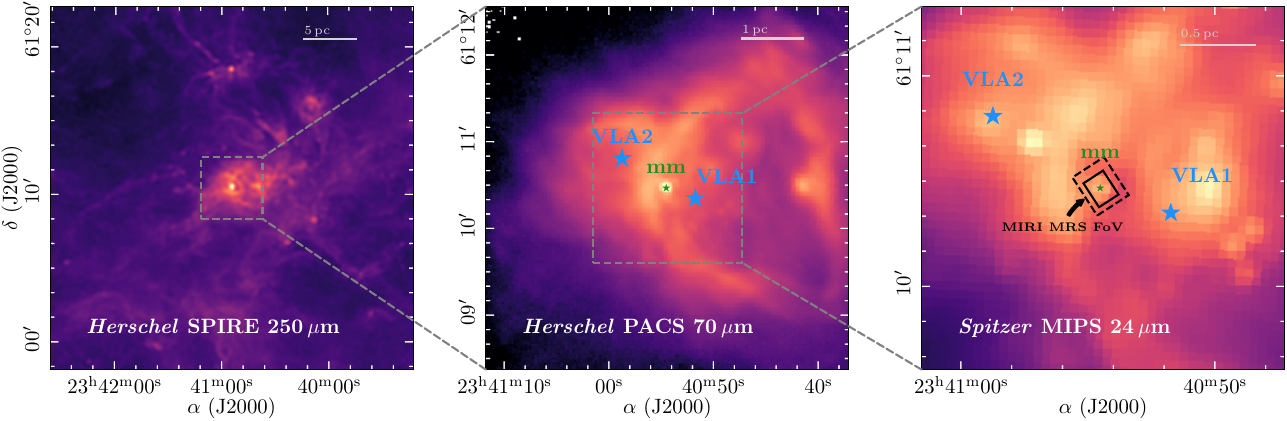}
\caption{Multi-wavelength overview of IRAS\,23385. In color, the \textit{Herschel} 250\,$\upmu$m (\textit{left}), \textit{Herschel} 70\,$\upmu$m (\textit{center}), and \textit{Spitzer} 24\,$\upmu$m (\textit{right}) emission are shown in log-scale. The grey dashed squares show the field-of-view of the following panel. In the center and right panels, the green star, labeled as ``mm'', marks the 1\,mm continuum peak position (Fig. \ref{fig:continuum}). The blue stars, labeled as ``VLA1'' and ``VLA2'', show the position of the two nearby UCH{\sc ii} regions \citep{Molinari2002}. The black rectangles in the right panel indicate the JWST/MIRI MRS field-of-view of the 4-pointing mosaic in ch1A (solid) and ch4C (dashed).}
\label{fig:overview}
\end{figure*}

	The formation of high-mass stars (stellar masses $M_\star > 8$\,$M_\odot$) still remains a puzzle, despite being studied for many decades, we refer to \citet{Motte2018} and \citet{Rosen2020} for recent reviews. High-mass star-forming regions (HMSFRs) are typically located at distances of several kiloparsecs and in order to achieve a spatial resolution of a few thousand au, sub-arcsecond resolution is necessary.
	
	
	With the recently launched \textit{James Webb Space Telescope} (JWST), infrared (IR) bright protostars can be characterized at sub-arcsecond resolution and high sensitivity in the near-infrared (NIR) and mid-infrared (MIR) regime. The observations analyzed in this work are part of the European Consortium guaranteed time program \textbf{J}WST \textbf{O}bservations of \textbf{Y}oung proto\textbf{S}tars (JOYS, PI: E. F. van Dishoeck, id. 1290). The project targets low-mass and high-mass protostars with the Mid-InfraRed Instrument (MIRI) using the Medium Resolution Spectrometer (MRS) mode. The first target observed as part of the JOYS project is the HMSFR IRAS\,23385+6053 (IRAS\,23385 hereafter), also referred to as ``Mol\,160'' \citep{Molinari1996}.
	
	IRAS\,23385 is associated with H$_{2}$O maser activity \citep{Molinari1996}, but remains undetected at cm wavelengths \citep{Molinari1998A}. This suggests that IRAS\,23385 is at an evolutionary stage prior to the ultra-compact (UC) H{\sc ii} region phase \citep{Molinari1998B}. The mass of the central object and the envelope are estimated to be $\approx$9\,$M_\odot$ \citep{Cesaroni2019} and 510\,$M_\odot$ \citep{Beuther2018}, respectively. \citet{Molinari1998B} derived a mean visual extinction of the core of $A_{V} = 200$\,mag with a H$_{2}$ column density $N$(H$_{2}$) of $2 \times 10^{24}$\,cm$^{-2}$.
	 
	
	The Two Micron All Sky Survey (2MASS) data toward the region reveal that the dense core is surrounded by NIR sources and is located in between two UCH{\sc ii} regions with bright cm emission \citep{Molinari2002,Thompson2003}. A large-scale overview toward the region is shown in Fig. \ref{fig:overview} highlighting the 250\,$\upmu$m, 70\,$\upmu$m, and 24\,$\upmu$m images in the region taken with the \textit{Herschel} and \textit{Spitzer} space telescopes. The central region is labeled ``mm'', while the two nearby UCH{\sc ii} regions are labeled ``VLA1'' and ``VLA2'' \citep{Molinari2002}. Large-scale filamentary cold dust emission is revealed by the \textit{Herschel} 250\,$\upmu$m data. The \textit{Herschel} 70\,$\upmu$m and \textit{Spitzer} 24\,$\upmu$m data reveal an emission arc between the UCH{\sc ii} region VLA2 and the mm core.
	
	
	\citet{Molinari2008} detect faint 24\,$\upmu$m emission (with a flux density of $F_{24\upmu\mathrm{m}} = 0.14$\,Jy) toward the position of the central dense core (their source ``A''), but no emission is detected at wavelengths below confirming the pre-ZAMS (zero-age main sequence) nature of the source. By modeling the spectral energy distribution (SED) the authors derive a bolometric luminosity of $L = 3.2 \times 10^3$\,$L_\odot$ for the core, while the luminosity of the entire region, i.e. including the two nearby bright UCH{\sc ii} regions (Fig. \ref{fig:overview}), is $L = 1.6 \times 10^4$\,$L_\odot$ \citep{Molinari1998B}.
	
	Bipolar outflows are found in broad line wings in CO emission \citep[southwest (redshifted)-northeast (blueshifted),][]{Wu2005} and in SiO and HCO$^{+}$ emission \citep[north (redshifted)-south (blueshifted),][]{Molinari1998B}. Since the emission is not extended and the outflow lobes overlap, \citet{Molinari1998B} conclude that the direction of the north-south outflow is nearly pole-on.
	
	IRAS\,23385 is part of the NOrthern Extended Millimeter Array (NOEMA) large program ``CORE'' (PI: H. Beuther) which aims at imaging the central dense core at an angular resolution of 0\farcs5 at 1\,mm \citep{Beuther2018}. \citet{Cesaroni2019} carried out a case-study toward IRAS\,23385 using the CORE data and suggested the presence of a northwest (redshifted)-southeast (blueshifted) outflow. These authors, based on the observed IR, mm, and cm emission, concluded that a cluster of massive stars is probably being formed in the central core of IRAS\,23385.
	
	
	While interferometric mm observations trace the cold material, the presence of outflows and MIR emission suggests an additional warm component that is heated by the protostars and/or shocks. As a homonuclear diatomic, the most abundant molecule in the interstellar medium, H$_{2}$, has no permanent dipole and therefore lacks a pure rotational dipole transitions. However, ro-vibrational quadrupole transitions of H$_{2}$ can be excited at high temperatures, with upper energy levels $E_\mathrm{u}$/$k_\mathrm{B} > 500$\,K. Such transitions can be accessed with JWST/MIRI for H$_{2}$ that covers the $\Delta J =0$ rotational lines in the vibrational ground state 0-0 from S(1) to S(7) (Table \ref{tab:MIRI_line_obs}). In addition, JWST/MIRI covers emission lines from neutral and ionized atoms and molecules. 
	
	The first JOYS results of IRAS\,23385 using JWST/MIRI MRS data are presented in \citet{Beuther2023}. Toward the central dense core, two MIR sources are detected at short wavelengths, that become unresolved at longer wavelengths ($\lambda \gtrsim 15$\,$\upmu$m). Aside from continuum emission, the MIRI spectrum from 5 to 28\,$\upmu$m is dominated by emission of H$_{2}$ lines and forbidden transitions of atomic lines tracing the MIR sources and multiple outflows. In addition, the spectrum shows deep absorption features by ice species. The continuum emission at 24\,$\upmu$m rises to $\approx$0.1-0.2\,Jy, a similar value derived from past \textit{Spitzer} data \citep[0.14\,Jy,][]{Molinari2008}. Using a faint Humphreys $\alpha$ line, an accretion rate of 0.9$\times$10$^{-4}$\,$M_\odot$\,yr$^{-1}$ is estimated \citep{Beuther2023}. Compact molecular line emission toward the MIR sources is being studied by Francis et al. (in prep.) and the ice properties will be presented in Rocha et al. (in prep.), see also \citet{vanDishoeck2023}.
	
	In this work, we characterize and compare in detail, for the first time, the warm and cold material, taking advantage of new JWST/MIRI MRS observations that can be compared with NOEMA data at mm wavelengths and to study the nature of the continuum sources. With NOEMA the cold molecular gas and dust properties can be studied, while with JWST/MIRI observations we probe the warmer environment. In Sect. \ref{sec:observations} the observational parameters and calibration of the JWST and NOEMA data are explained. The analysis of the continuum and spectral line data is presented in Sect. \ref{sec:results} and in Sect. \ref{sec:discussion} the results are discussed. Our conclusions are summarized in Sect. \ref{sec:conclusions}.
	
\section{Observations}\label{sec:observations}

	In this work we analyze line and continuum observations at MIR wavelengths with JWST/MIRI (Sect. \ref{sec:JWSTobs}) and at 1\,mm and 3\,mm with NOEMA observations (Sect. \ref{sec:NOEMAobs}). With angular resolutions $0\farcs2 - 1''$, spatial scales down to $\approx$1\,000\,au-5\,000\,au can be traced in IRAS\,23385, at a distance of 4.9\,kpc.
	
	The kinematic distance is estimated to be 4.9\,kpc \citep{Molinari1998B} with a Local Standard of Rest (LSR) velocity of $\varv_\mathrm{LSR} = -50.2$\,km\,s$^{-1}$ \citep{Beuther2018}. Typically, kinematic distance estimates suffer from high uncertainties. \citet{Molinari2008} modeled the SED with a zero-age main-sequence (ZAMS) star embedded in an envelope but were unsuccessful in explaining the observed SED properly. The authors argued that the region may actually be located at 8\,kpc or that the source is at a younger evolutionary stage. On the other hand, using the Galactic rotation curve and spiral arm model by \citet{Reid2016,Reid2019}, a distance of 2.7$\pm$0.2\,kpc could be estimated\footnote{\url{http://bessel.vlbi-astrometry.org/node/378}}. However, this distance estimate includes as a prior its association with a spiral arm which is unknown. Maser parallax measurements are not available for this source. \citet{Choi2014} measured maser parallaxes toward high-mass star-forming regions and compared the corresponding distances with kinematic distance estimates finding significant discrepancies. The authors concluded that for G111.23$-$1.23, the source in their sample that is closest to IRAS\,23385 and has a similar $\varv_\mathrm{LSR}$, the kinematic distance is 4.8\,kpc ($\varv_\mathrm{LSR} = -53$\,km\,s$^{-1}$), while the maser parallax measurement suggests a distance of 3.33\,kpc. However, since IRAS23385 is not covered by \citet{Choi2014}, and given the uncertainties, we assume in this work a distance of 4.9\,kpc, consistent with previous studies in the literature and in line with the distance used by the JOYS team \citep{Beuther2023}. The high uncertainty in distance does not affect the analysis in this work.

\subsection{JWST}\label{sec:JWSTobs}

	The observations were carried out on August 22, 2022 with a total observing time of 2.12\,h (40\,min on source) with a mosaic of four pointings surrounding the dense core in IRAS\,23385. The MIRI MRS instrument consists of four channels (ch1, ch2, ch3, ch4) and each of them is divided intro three sub-bands (short, medium, and long) to which we refer as ``A'', ``B'', and ``C''. The total spectral coverage ranges from 4.9\,$\upmu$m to 27.9\,$\upmu$m with a resolving power $R$ decreasing from 3\,700 to 1\,300 \citep[][]{Wells2015, Labiano2021,Argyriou2023, Jones2023}. The angular resolution can be estimated as $\theta = 0.033 \lambda (\upmu\mathrm{m}) + 0\farcs106$ \citep{Law2023} and field-of-view (FoV) of the 4-pointing mosaic increases from $\approx 7''\times 7''$ to $\approx 15''\times 15''$ from short to long wavelengths (Fig. \ref{fig:overview}). The slice width of MIRI MRS increases from 0\farcs176 to 0\farcs645 from ch1 to ch4 \citep{Wells2015}.
	
	The observations were calibrated using the JWST Science Calibration Pipeline (version 1.11.2) and Calibration Reference Data System (CRDS) context ``jwst\_1100.pmap''. In addition, astrometric corrections were applied by comparing detected sources in the MIRI imaging field, that was observed simultaneously with MIRI MRS, with Gaia data \citep{GaiaEDR3}. We refer to \citet{Beuther2023} for a detailed description of the calibration of this data set.
	
	The resulting spectral cube data product consists of continuum and line emission, as well as ice absorption features. In this work, we focus on the extended H$_{2}$ emission and forbidden atomic fine structure transitions. In Table \ref{tab:MIRI_line_obs} we give an overview of all lines that are analyzed in this work. With a spectral resolution of a few tens of km\,s$^{-1}$, the high-velocity outflow components can be resolved \citep{Beuther2023}, but a detailed kinematic analysis of the gas properties close to the protostars is difficult. Thus, in this work we focus on the spatial distribution of the integrated emission.
	
\setlength{\tabcolsep}{1pt}
\begin{table}[!htb]
\caption{MIR lines covered by JWST/MIRI MRS.}
\label{tab:MIRI_line_obs}
\centering
\begin{tabular}{lrrrrr}
\hline\hline
Line & Wavelength & Upper & Angular &  Line & MIRI \\
 & & energy & reso- & noise & MRS \\
 & & level & lution & & unit \\
 & $\lambda$ & $E_\mathrm{u}$/$k_\mathrm{B}$ & $\theta$ & $\sigma_\mathrm{line}$ & \\
 & ($\upmu$m) & (K) & ($''$) & (MJy\,sr$^{-1}$) & \\
\hline
H$_{2}$\,0-0\,S(7) & 5.511 & 7\,197 &  0.29 & 38 & ch1A\\ 
H$_{2}$\,0-0\,S(6) & 6.109 & 5\,829 &  0.31 & 34 & ch1B\\ 
$[$Ni\,\sc{ii}$]$\,$^{2}D_{3/2}$-$^{2}D_{5/2}$ & 6.636 & 2\,168 &  0.32 & 27 & ch1C\\ 
H$_{2}$\,0-0\,S(5) & 6.910 & 4\,586 &  0.33 & 27 & ch1C\\ 
$[$Ar\,\sc{ii}$]$\,$^{2}P_{1/2}$-$^{2}P_{3/2}$ & 6.985 & 2\,060 &  0.34 & 27 & ch1C\\ 
H$_{2}$\,0-0\,S(4) & 8.025 & 3\,474 &  0.37 & 19 & ch2A\\ 
H$_{2}$\,0-0\,S(3) & 9.665 & 2\,504 &  0.42 & 18 & ch2B\\ 
H$_{2}$\,0-0\,S(2) & 12.279 & 1\,682 &  0.51 & 7.2 & ch3A\\ 
$[$Ne\,\sc{ii}$]$\,$^{2}P^{0}_{1/2}$-$^{2}P^{0}_{3/2}$ & 12.814 & 1\,123 &  0.53 & 7.2 & ch3A\\ 
$[$Ne\,\sc{iii}$]$\,$^{3}P_{1}$-$^{3}P_{2}$ & 15.555 & 925 &  0.62 & 5.2 & ch3C\\ 
H$_{2}$\,0-0\,S(1) & 17.035 & 1\,015 &  0.67 & 7.3 & ch3C\\ 
$[$Fe\,\sc{ii}$]$\,$^{4}F_{7/2}$-$^{4}F_{9/2}$ & 17.936 & 3\,496 &  0.70 & 8.7 & ch3C\\ 
$[$S\,\sc{i}$]$\,$^{3}P_{1}$-$^{3}P_{2}$ & 25.249 & 570 &  0.94 & 32 & ch4C\\ 
\hline
\end{tabular}
\tablefoot{The wavelength and upper energy level of the H$_{2}$ lines are taken from \citet{Jennings1987} and for the remaining transitions from the atomic line list by \citet{vanHoof2018}.}
\end{table}
	
\setlength{\tabcolsep}{5pt}
\begin{table*}[!htb]
\caption{Molecular lines covered by NOEMA at 1\,mm and 3\,mm analyzed in this work.}
\label{tab:mm_line_obs}
\centering
\begin{tabular}{lrrrrrrr}
\hline\hline
 & & Upper & & & Channel & & NOEMA \\
Line & Frequency & energy level & \multicolumn{2}{c}{Synthesized Beam} & width & Line noise & project \\ \cline{4-5}
 & $\nu$ & $E_\mathrm{u}$/$k_\mathrm{B}$ & $\theta_\mathrm{maj}\times\theta_\mathrm{min}$ & PA & $\delta \varv$ & $\sigma_\mathrm{line}$ & \\
 & (GHz) & (K) & ($'' \times ''$) & ($^\circ$) & (km\,s$^{-1}$) & (K) & \\
\hline
H$^{13}$CO$^{+}$ ($1-0$) & 86.754 & 4.2 & 1.43$\times$1.18 & 50.3 & 0.8 & 0.18 & CORE+\\ 
SiO ($2-1$) & 86.847 & 6.3 & 1.43$\times$1.18 & 50.3 & 0.8 & 0.18 & CORE+\\ 
HC$_{3}$N ($10-9$) & 90.979 & 24.0 & 1.39$\times$1.13 & 52.3 & 0.8 & 0.16 & CORE+\\ 
CH$_{3}$CN ($5_{3}-4_{3}$) & 91.971 & 77.5 & 1.37$\times$1.12 & 52.6 & 0.8 & 0.16 & CORE+\\ 
CH$_{3}$CN ($5_{2}-4_{2}$) & 91.980 & 41.8 & 1.37$\times$1.12 & 52.6 & 0.8 & 0.19 & CORE+\\ 
CH$_{3}$CN ($5_{1}-4_{1}$) & 91.985 & 20.4 & 1.37$\times$1.12 & 52.6 & 0.8 & 0.16 & CORE+\\ 
CH$_{3}$CN ($5_{0}-4_{0}$) & 91.987 & 13.2 & 1.37$\times$1.12 & 52.6 & 0.8 & 0.17 & CORE+\\ 
$^{13}$CS ($2-1$) & 92.494 & 6.7 & 1.37$\times$1.12 & 52.8 & 0.8 & 0.17 & CORE+\\ 
CH$_{3}$OH ($4_{2,3}-3_{1,2}E$) & 218.440 & 45.5 & 0.45$\times$0.44 & 55.2 & 0.5 & 1.20 & CORE\\ 
OCS ($18-17$) & 218.903 & 99.8 & 0.45$\times$0.44 & 15.5 & 0.5 & 1.21 & CORE\\ 
SO ($6_{5}-5_{4}$) & 219.949 & 35.0 & 0.49$\times$0.44 & 54.8 & 3.0 & 0.40 & CORE\\ 
H$_{2}$CO ($3_{0,3}-2_{0,2}$) & 218.222 & 21.0 & 0.49$\times$0.44 & 54.8 & 3.0 & 0.38 & CORE\\ 
\hline 
\end{tabular}
\tablefoot{The 1\,mm and 3\,mm data were taken as part of the CORE \citep{Beuther2018} and CORE+ (Gieser et al., in prep.) programs, respectively.}
\end{table*}

	A 5.2\,$\upmu$m continuum map is created using the ch1A band at a wavelength range between 5.2\,$\upmu$m and 5.3\,$\upmu$m in which bright line emission or absorption features are absent. The angular resolution is $\approx$0\farcs2. Faint extended background emission with a median of 48\,MJy\,sr$^{-1}$ is subtracted from the 5.2\,$\upmu$m intensity map. The noise measured as the standard deviation of the 5.2\,$\upmu$m continuum map is 18\,MJy\,sr$^{-1}$ after background subtraction.
	
	In individual line cubes (Table \ref{tab:MIRI_line_obs}), the local continuum was subtracted by fitting a first order polynomial to the spectra masking out channels with line emission. The line noise, $\sigma_\mathrm{line}$, is estimated in line-free channels and decreases from 38\,MJy\,sr$^{-1}$ to 5\,MJy\,sr$^{-1}$ from ch1A to ch3C in which the noise increases again up to ch4C to 32\,MJy\,sr$^{-1}$.

\subsection{NOEMA}\label{sec:NOEMAobs}

	Located in the Northern hemisphere, with a high declination of $\approx$61$^\circ$, IRAS\,23385 is accessible best at high angular resolution at mm wavelengths with the NOrthern Extended Millimeter Array (NOEMA).

	\subsubsection{1 mm data}

	IRAS\,23385 is included in the sample of the NOEMA large program CORE \citep[][PI: H. Beuther, project code L14AB]{Beuther2018, Gieser2021, Ahmadi2023}. The region was observed with NOEMA in the D, A, and C configurations in January 2015, February 2016, and October 2016, respectively. The WideX correlator provides a spectral coverage from 217.2\,GHz to 220.8\,GHz (1.36\,mm-1.38\,mm) at an angular resolution of $\approx$0\farcs5. High resolution data with a channel width, $\delta \varv$, of 0.5\,km\,s$^{-1}$ are available for a few targeted molecular lines \citep[Table 2 and 3 in][]{Ahmadi2018}. A continuous spectrum along the full bandwidth is available with a channel width of 3.0\,km\,s$^{-1}$. For the spectral line data, complementary short spacing observations with the IRAM\,30m telescope were obtained recovering missing flux filtered out by the interferometer. The combination of the interferometric and single-dish (``merged'') data is explained in \citet{Mottram2020}.
	
	The NOEMA data were calibrated using the {CLIC} package in \texttt{GILDAS}\footnote{\url{https://www.iram.fr/IRAMFR/GILDAS/}}. The interferometric line and continuum data were self-calibrated in order to increase the signal-to-noise ($S$/$N$) ratio. Phase self-calibration was performed on the continuum data and then applied to the spectral line data. A detailed description of the self-calibration procedure for the NOEMA data can be found in \citet{Gieser2021}.
	 
	 The NOEMA continuum and continuum-subtracted merged spectral line data were deconvolved using the \texttt{GILDAS/MAPPING} package and the Clark algorithm. The weighting was set to a robust parameter of 0.1 to achieve the highest possible angular resolution. For the 1\,mm continuum data, the resulting synthesized beam (with major and minor axis $\theta_\mathrm{maj}\times\theta_\mathrm{min}$) is 0\farcs48$\times$0\farcs43 with a position angle (PA) of 58$^\circ$. The continuum noise is estimated to be $\sigma_\mathrm{cont,1mm}$ = 0.15\,mJy\,beam$^{-1}$. 
	 
	 The properties of the 1\,mm molecular lines presented in this work are summarized in Table \ref{tab:mm_line_obs} with the label "CORE" in the NOEMA project column. The line sensitivity $\sigma_\mathrm{line}$, estimated in channels without line emission, is $\approx$1.2\,K and $\approx$0.4\,K in the high- and low spectral resolution data, respectively, at an angular resolution of $\approx$0\farcs5.
	 
	 \subsubsection{3 mm data}
	 
	 Using the upgraded PolyFiX correlator at NOEMA, CORE+ (PI: C. Gieser, project code W20AV) is a follow-up project of CORE targeting the 3\,mm wavelength range. In a first case-study, three regions of the CORE sample, including IRAS\,23385, were selected (Gieser et al., in prep.). The NOEMA 3\,mm observations were obtained in the A and D configuration and short-spacing information was provided by complementary IRAM 30\,m observations. The observations were carried out from February to April 2021 with NOEMA and in August 2021 with the IRAM 30m telescope. Using the high-resolution spectral units, a channel width $\delta \varv$ of 0.8\,km\,s$^{-1}$ is achieved in the merged line data.
	 
	 Phase self-calibration was performed on the continuum data using the same procedure as in the 1\,mm CORE data \citep{Gieser2021} and the gain solutions were applied to the NOEMA line data. The NOEMA continuum and continuum-subtracted merged spectral line data were deconvolved using the \texttt{GILDAS/MAPPING} package and the Clark algorithm. The weighting was set to a robust parameter of 1 to achieve a compromise between high angular resolution and good sensitivity. For the 3\,mm continuum data, the resulting synthesized beam is 1\farcs1$\times$0\farcs85 (PA 47$^\circ$) and the noise is estimated to be $\sigma_\mathrm{cont,3mm}$ = 0.020\,mJy\,beam$^{-1}$. 
	 
\begin{figure*}[!htb]
\centering
\includegraphics[width=0.475\textwidth]{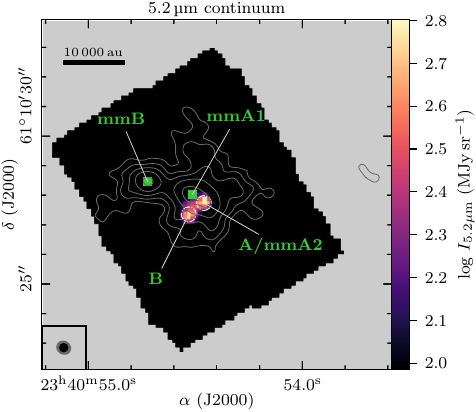}
\includegraphics[width=0.495\textwidth]{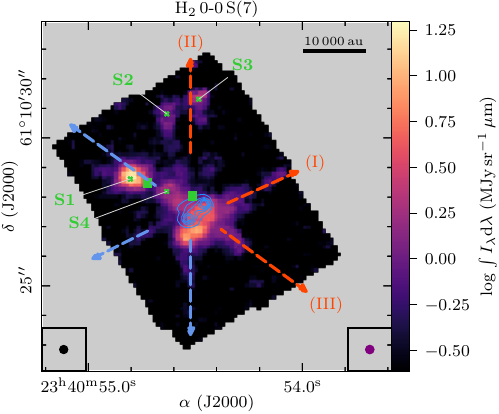}
\caption{MIR Continuum (\textit{left}) and H$_{2}$ 0-0 S(7) line emission (\textit{right}) for IRAS\,23385. In the left panel, the JWST/MIRI 5.2\,$\upmu$m continuum with emission $> 5\sigma_{\mathrm{cont,5}\upmu\mathrm{m}}$ is presented in color. Grey contours show the NOEMA 1\,mm continuum with contour levels at 5, 10, 20, 40, 80$\times \sigma_\mathrm{cont,1mm}$. The dash-dotted white circles show the aperture in which the 5.2\,$\upmu$m flux density $F_{5.2\upmu\mathrm{m}}$ was derived (Table \ref{tab:sources}). In the bottom left corner, the JWST/MIRI 5.2\,$\upmu$m and NOEMA 1\,mm angular resolution is indicated in black and grey, respectively. The mm (mmA1 and mmB, marked by green squares) and MIR (A/mmA2 and B) continuum sources are labeled in green. In the right panel, the line integrated intensity of the H$_{2}$ 0-0 S(7) line with $S$/$N > 5$ is presented in color. The JWST/MIRI 5.2\,$\upmu$m continuum as shown in the left panel is presented as blue contours with contour levels at 5, 10, 15, 20, 25$\times \sigma_{\mathrm{cont,5}\upmu\mathrm{m}}$. The angular resolution of the 5.2\,$\upmu$m continuum and H$_{2}$ 0-0 S(7) line data is indicated in the bottom left and right, respectively. Several shock spots evident in the MIRI MRS line emission are marked by green crosses and labeled in green (Sect. \ref{sec:line}). The red and blue arrows indicate three bipolar outflows, labeled I, II, and III \citep[as presented in][]{Beuther2023}. In both panels, the black bar indicates a spatial scale of 10\,000\,au at the assumed source distance of 4.9\,kpc.}
\label{fig:continuum}
\end{figure*}	 
	 The properties of the 3\,mm molecular lines analyzed in this study are summarized in Table \ref{tab:mm_line_obs} with the label ``CORE+'' in the last column. The line sensitivity $\sigma_\mathrm{line}$ at a channel width of 0.8\,km\,s$^{-1}$ is $\approx0.2$\,K.
	 
	 For all 1\,mm and 3\,mm NOEMA continuum and line data products we apply a primary beam correction taking into account that the noise levels increase toward the edge of the primary beam. The primary beam is $\approx$22$''$ and $\approx$60$''$ at 1\,mm and 3\,mm, respectively. Thus, a larger FoV is covered in the NOEMA data compared to the 4-pointing mosaic JWST/MIRI observations ($\approx15''$ at the longest wavelengths). Since we aim to compare the MIR and mm data, we only focus here on the central area ($\approx 12''\times12''$) that is also covered by the JWST/MIRI FoV. The upper energy levels $E_\mathrm{u}$/$k_\mathrm{B}$ of the mm lines vary between 4\,K and 100\,K, that is significantly lower than for the JWST/MIRI transitions ($> 500$\,K). Thus, with the NOEMA continuum and line data, we are able to trace in the same region also the cold molecular gas.

\section{Results}\label{sec:results}

	A comparison of the detected JWST/MIRI MRS and NOEMA mm lines allows us to study the properties of the warm and cold material, respectively, at sub-arcsecond resolution corresponding to spatial scales below 5\,000\,au for IRAS\,23385. In addition, the continuum allows us to map the different sources present in the FoV. In Sect. \ref{sec:continuum} the continuum emission, detected sources, and outflows are presented. The spatial morphology of atomic and molecular line emission is compared in Sect. \ref{sec:line}. The temperature of the warmer gas ($>$100\,K) is derived from the JWST/MIRI H$_{2}$ 0-0 lines in Sect. \ref{sec:H2_modeling}, while the temperature of the cold gas ($<$100\,K) is estimated by modeling CH$_{3}$CN emission lines in Sect. \ref{sec:CH3CN_modeling}.

\subsection{Continuum sources and outflows}\label{sec:continuum}

	A comparison of the JWST/MIRI 5.2\,$\upmu$m and NOEMA 1\,mm continuum data is presented in Fig. \ref{fig:continuum} in the left panel with detected sources labeled. For the mm and MIR sources, we follow the nomenclature of \citet{Cesaroni2019} and \citet{Beuther2023}, respectively. 
	
	The 1\,mm continuum data clearly resolve two cores (mmA1 and mmB) that are separated by $\approx$1\farcs6, corresponding to a projected separation of $\approx$7\,700\,au. The mmA continuum peak is elongated, can be separated into two sources ``mmA1'' and ``mmA2''.
	
	The 5.2\,$\upmu$m continuum data reveal two sources, A and B, that are barely resolved at an angular resolution of 0\farcs2. The MIR source A is co-spatial with the elongated mm core ``mmA2'', hereafter referred to as ``A/mmA2''. While in the shorter wavelength regime of MIRI the two sources can be resolved, beyond $\approx$15\,$\upmu$m this is not possible due to the lower angular resolution \citep{Beuther2023}. Notably, no additional source is detected in the MIRI data at longer wavelengths. The positions of the four continuum sources are summarized in Table \ref{tab:sources}. In addition, the flux density at 5.2\,$\upmu$m $F_{5.2\upmu\mathrm{m}}$, calculated within a 0\farcs5 circular aperture (Fig. \ref{fig:continuum}), of the MIR sources A/mmA2 and B are listed. For the non-detections of sources mmA1 and mmB, $3\sigma$ upper limits are given.
	
\setlength{\tabcolsep}{5pt}
\begin{table}[!htb]
\caption{Continuum sources in IRAS\,23385.}
\label{tab:sources}
\centering
\begin{tabular}{lccr}
\hline\hline
Source & \multicolumn{2}{c}{Position} & Flux density \\ \cline{2-3}
 & $\alpha$ & $\delta$ & $F_{5.2\upmu\mathrm{m}}$\\
 & \multicolumn{2}{c}{J(2000)} & (mJy)\\
\hline
mmA1 & 23:40:54.51 & $+$61:10:28.0 & $<$0.023\\ 
mmB & 23:40:54.72 & $+$61:10:28.5 & $<$0.023\\ 
A/mmA2 & 23:40:54.46 & $+$61:10:27.7 & 1.641\\ 
B & 23:40:54.53 & $+$61:10:27.3 & 1.594\\ 
\hline 
\end{tabular}
\tablefoot{The flux density $F_{5.2\upmu\mathrm{m}}$ is computed within a 0\farcs5 aperture (Fig. \ref{fig:continuum}). In case of non-detections, 3$\sigma$ upper limits are listed.}
\end{table}
	
	The 1\,mm continuum morphology surrounding the cores reveals complex substructure. The mm continuum emission is elongated toward the north from position mmA1, but also elongated toward the southeast from position mmB. The 1\,mm (and also the 3\,mm) continuum data trace cold dust emission that is optically thin: \citet{Gieser2021} estimate that toward mmA1 and mmB the values of the continuum optical depth at 1\,mm, $\tau_\mathrm{cont,1mm}$, are 9.3$\times$10$^{-3}$ and 7.9$\times$10$^{-3}$, respectively. In the 3\,mm continuum data, we also find $\tau_\mathrm{cont,3mm} \ll 1$ (Sect. \ref{sec:CH3CN_modeling}). The core masses of mmA1 and mmB are estimated to be 21.9\,$M_\odot$ and 9.4\,$M_\odot$ \citep[with $T_\mathrm{kin} = 73$\,K,][]{Beuther2018}. Thus, mmA1 contains about twice the mass compared to mmB. These mass estimates are lower limits due to spatial filtering of the extended emission in the interferometric observations.

	In summary, the NOEMA observations trace the optically thin dust emission of a cold envelope in which the four sources are embedded. The two MIR sources reveal two more evolved protostars with a hot thermal component. While source A has a clear counterpart at mm wavelengths (mmA2), source B does not coincide with a fragmented mm core, which may be due to a poorer angular resolution (0\farcs5) of the NOEMA data compared to the JWST/MIRI 5.2\,$\upmu$m data (0\farcs2). Still, source B is embedded within the cold dust envelope. The fact that multiple sources are detected toward the central region of IRAS\,23385 reinforces the idea that a cluster is forming \citep{Cesaroni2019}. The sources mmA1 and mmB with no MIR-bright counterpart could be either younger compared to A/mmA2 and B or their MIR-extinction is high enough to completely block the protostellar MIR emission. The nature of the continuum sources is further discussed in Sect. \ref{sec:discussion}.
	
	The right panel in Fig. \ref{fig:continuum} shows the integrated intensity of the H$_{2}$ 0-0 S(7) line at 5.511\,$\upmu$m (Table \ref{tab:MIRI_line_obs}), showing H$_{2}$ emission at high angular resolution, revealing the presence of multiple bipolar outflows. Following \citet{Beuther2023}, the outflows I, II, and III could be associated with the continuum sources B, mmA1, and A/mmA2, respectively. We note that due to the clustered nature of IRAS\,23385, the outflow driving sources cannot be identified unambiguously. We refer here to the \textit{most likely} driving source, as discussed in \citet{Beuther2023}. Complementary NIR observations with JWST that have even higher angular resolution could provide a more reliable assignment. The outflow directions are indicated by arrows in Fig. \ref{fig:continuum}. In addition, several positions where strong shocks occur are marked by green crosses and are labeled as S1, S2, S3, and S4. In addition, bright H$_{2}$ emission is detected toward the south of source B and the southwest of source A/mmA2 tracing shocks and heated material very close to the protostars. While these shocked locations are already evident in the H$_{2}$ 0-0 S(7) map, we find that additional forbidden transitions of atomic lines can be present there, as discussed in the following section. No outflow could be linked to source mmB, since the close-by bright H$_{2}$ knot (S1) toward the northeast of mmB is most likely connected to outflow III \citep{Beuther2023}.

\subsection{Atomic and molecular line emission}\label{sec:line}
\begin{figure*}[!htb]
\centering
\includegraphics[width=0.33\textwidth]{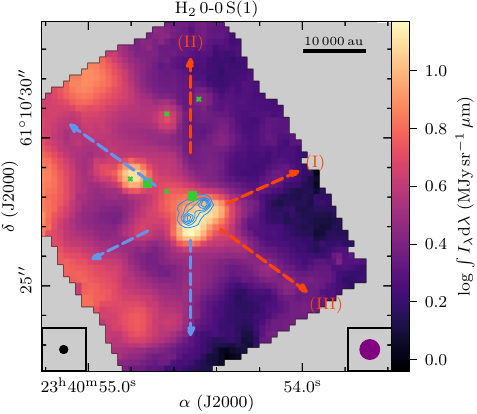}
\includegraphics[width=0.33\textwidth]{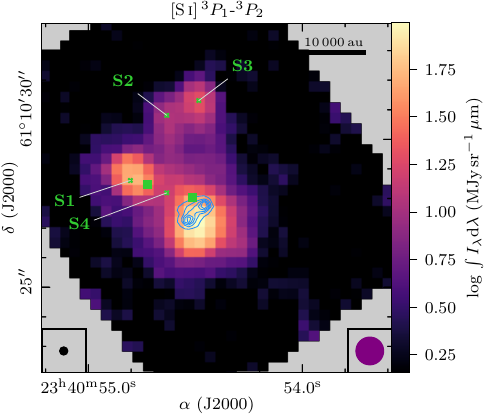}\\
\includegraphics[width=0.33\textwidth]{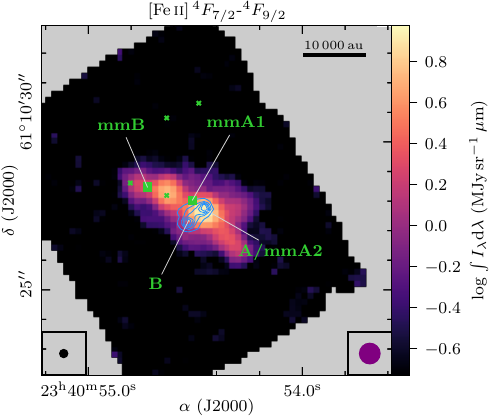}
\includegraphics[width=0.33\textwidth]{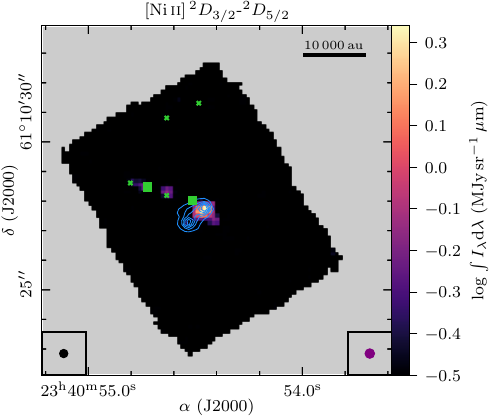}\\
\includegraphics[width=0.33\textwidth]{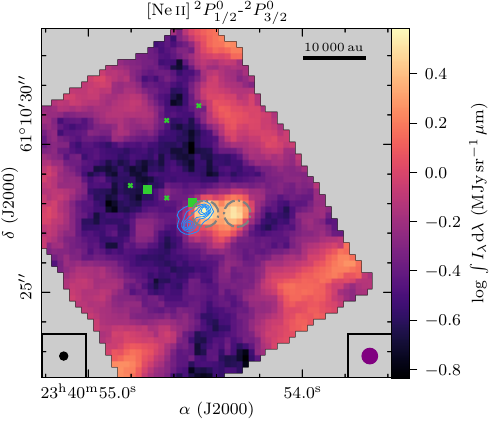}
\includegraphics[width=0.33\textwidth]{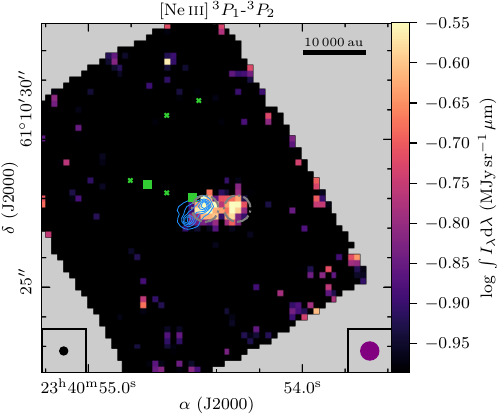}
\includegraphics[width=0.33\textwidth]{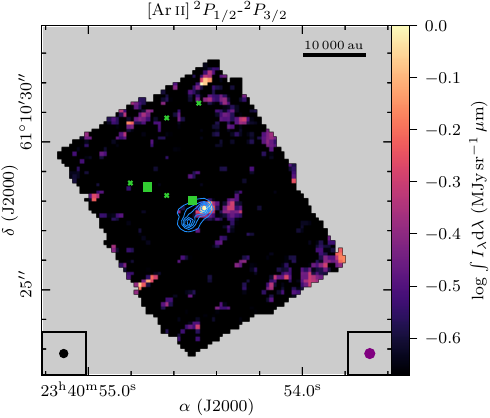}
\caption{Integrated intensity maps of atomic and molecular lines detected with JWST/MIRI MRS. In color, the line integrated intensity is shown in a log-scale. The JWST/MIRI 5.2\,$\upmu$m continuum is presented as blue contours with contour levels at 5, 10, 15, 20, 25$\times \sigma_{\mathrm{cont,5}\upmu\mathrm{m}}$. The two mm sources are indicated by green squares and several shock positions are highlighted by green crosses. In the bottom left and right corners, the angular resolution of the JWST/MIRI continuum and line data, respectively, is shown. In the top left panel, the bipolar outflows are indicated by red and blue dashed arrows. The shock locations (Sect. \ref{sec:line}) and continuum sources are labeled in the top right and center left panel, respectively. In the $[$Ne\,{\sc ii}$]$ and $[$Ne\,{\sc iii}$]$ panels, the dash-dotted grey circles show the aperture (0\farcs9) in which the flux density was derived (Sect. \ref{sec:discussion}).}
\label{fig:line}
\end{figure*}
	In this work, we focus on the spatial distribution of H$_{2}$ and the neutral and ionized atomic forbidden fine-structure lines, while a detailed analysis of the MIR molecular emission lines other than H$_{2}$ covered by JWST/MIRI, for example CO, CO$_{2}$, and HCN, will be presented in Francis et al. (in prep.). The H$_{2}$ transitions covered by JWST/MIRI only trace the warmer/hotter gas (Table \ref{tab:MIRI_line_obs}, Sect. \ref{sec:H2_modeling}), typically the heated and/or shocked gas along the jet or gas entrained by the jet. Therefore, we use CH$_{3}$CN and the mm dust emission to trace the cold H$_{2}$ component (Sect. \ref{sec:CH3CN_modeling}). In this section, we first compare the spatial morphology of the emission of atoms and molecules detected by JWST and NOEMA.
	
	All studied transitions covered by JWST/MIRI are listed in Table \ref{tab:MIRI_line_obs}, but in the following, for H$_{2}$ we only present the spatial morphology of the 0-0 S(1) transition, while for the S(7) transition it is shown in the right panel in Fig. \ref{fig:continuum}, since the spatial morphology of the remaining H$_{2}$ lines is similar. The studied molecular transitions covered by NOEMA are summarized in Table \ref{tab:mm_line_obs}. In order to obtain the spatial distribution of each emission line, we compute the line integrated intensity of the continuum-subtracted data cubes, $\int_{\lambda_\mathrm{low}}^{\lambda_\mathrm{upp}} I_\lambda \mathrm{d}\lambda$ and $\int_{\varv_\mathrm{low}}^{\varv_\mathrm{upp}} I_\varv \mathrm{d}\varv$, for the JWST and NOEMA observations, respectively. Integration ranges were selected based on visual inspection of the emission. The integrated intensity maps of the emission lines are presented in Figs. \ref{fig:line} (MIR lines) and \ref{fig:line_mm} (mm lines) with a comparison to the JWST/MIRI 5.2\,$\upmu$m continuum, shock positions, and outflow directions.

\subsubsection{Molecular hydrogen H$_{2}$}

	The 0-0 S(1) transition of H$_{2}$ is detected everywhere within the field-of-view (FoV) of MIRI revealing at least three outflows \citep{Beuther2023}. Bright emission peaks are found slightly toward the south and southwest of the MIR-bright sources B and A/mmA2, respectively. Additional bright H$_{2}$ emission peaks are found toward the east and north of mmB and we refer to these positions as ``S1'' and ``S2'' in the following and they are also marked in Fig. \ref{fig:line} (top right panel). Another interesting feature is a bright emission arc toward the east at the edge of the MIRI FoV. This arc-like structure is not seen in the S(7) integrated intensity map (Fig. \ref{fig:continuum}), most likely due to a smaller FoV in the MIRI MRS data. This is not an artifact, but is most likely caused by the irradiation of the nearby UCH{\sc ii} region heating up the environment and is also seen in the \textit{Spitzer} 24\,$\upmu$m image (Fig. \ref{fig:overview}). Otherwise, both H$_{2}$ transitions trace the same spatial structures in the dense core: heated and shocked gas along the outflow direction.
	
\subsubsection{Atomic sulfur $[$S\,{\sc i}$]$}

	Atomic sulfur, $[$S\,{\sc i}$]$, is typically associated with high-density material of 10$^{5}$\,cm$^{-3}$ due to a high critical density of the line in non-dissociative $C$-type shocks \citep{Hollenbach1989, Neufeld2007, Anderson2013}. The atomic sulfur $[$S\,{\sc i}$]$ emission line follows a similar morphology as H$_{2}$. The emission arc toward the east is not detected in $[$S\,{\sc i}$]$, however, another bright emission peak, ``S3'', toward the north of mmA1 and close to S2 has enhanced $[$S\,{\sc i}$]$ emission. This knot is also detected faintly in H$_{2}$ 0-0 S(1), but more clearly in H$_{2}$ 0-0 S(7) (Fig. \ref{fig:continuum}). Both the H$_{2}$ and $[$S\,{\sc i}$]$ emission reveal shocked gas toward the center of the IRAS\,23385 region caused by protostellar outflows. The north-south outflow (II) most likely originates from source mmA1 \citep{Beuther2023} with strong H$_{2}$ and $[$S\,{\sc i}$]$ emission close to the source toward the south and the shocked gas at the positions S2 and S3 toward the north.
	
\subsubsection{Iron $[$Fe\,{\sc ii}$]$ and Nickel $[$Ni\,{\sc ii}$]$}

	Fine-structure lines from ionized atoms, such as iron (Fe) and nickel (Ni), primarily arise in dissociative shock fronts in protostellar outflows \citep{Neufeld2009}. The $[$Fe\,{\sc ii}$]$ and $[$Ni\,{\sc ii}$]$ lines have a similar spatial distribution to each other, both peak toward source A/mmA2, with elongated emission from northeast to southwest. Along this elongation another emission peak, S4, appears in these lines that is located right between the two mm continuum peak positions. This suggests that source A/mmA2 drives an outflow from the northeast to southwest (III) and that S4 and S1 reveal shocked gas along the outflow. Interestingly, the emission is not continuous, but shows emission knots suggesting an outflow with strong episodic outbursts. Outflow III is the only one that shows emission by these ionized atoms.
	
\subsubsection{Neon ($[$Ne\,{\sc ii}$]$ and $[$Ne\,{\sc iii}$]$) and Argon $[$Ar\,{\sc ii}$]$}

	The spatial morphology of $[$Ne\,{\sc ii}$]$, $[$Ne\,{\sc iii}$]$, and $[$Ar\,{\sc ii}$]$ are similar, peaking toward the position of source A/mmA2 and with extended emission toward the west. Their spatial distribution is very different from that of $[$Fe\,{\sc ii}$]$ and $[$Ni\,{\sc ii}$]$. This indicates that toward this source, the internal UV and X-ray radiation is high enough to ionize these species and to excite such transitions with upper energy levels in the range of 1\,000\,K - 3\,500\,K (Table \ref{tab:MIRI_line_obs}). There is a secondary peak, 0\farcs9 toward the west of source A/mmA2, though only faintly seen in $[$Ar\,{\sc ii}$]$. This feature is further discussed in Sect. \ref{sec:discussion}.
	
	The $[$Ne\,{\sc ii}$]$ emission also shows an extended component with bright emission toward the edges of the FoV. While $[$Ne\,{\sc ii}$]$ can arise from $J$-type shocks \citep{Hollenbach1989} toward the protostars, the extended emission arises from energetic radiation of the nearby UCH{\sc ii} regions (Fig. \ref{fig:overview}).
	
\subsubsection{Millimeter lines}
\begin{figure*}[!htb]
\centering
\includegraphics[width=0.33\textwidth]{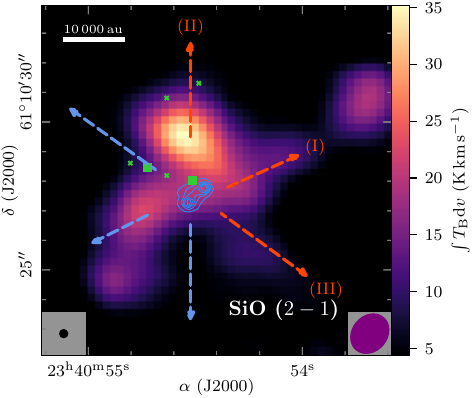}
\includegraphics[width=0.33\textwidth]{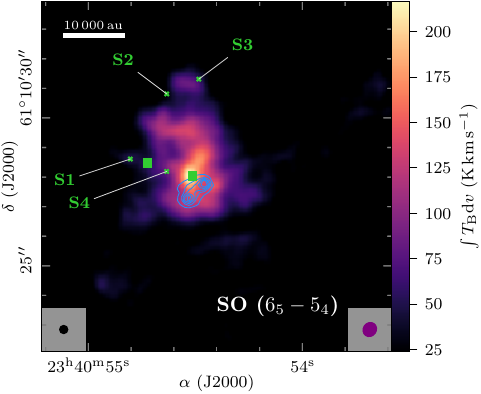}
\includegraphics[width=0.33\textwidth]{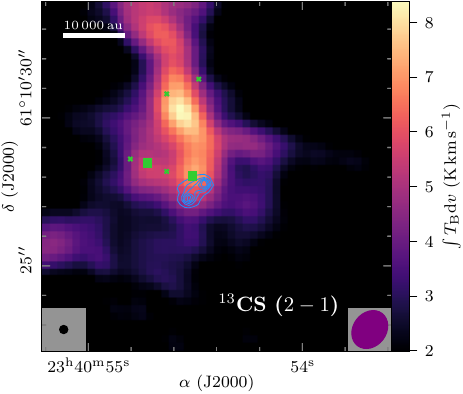}
\includegraphics[width=0.33\textwidth]{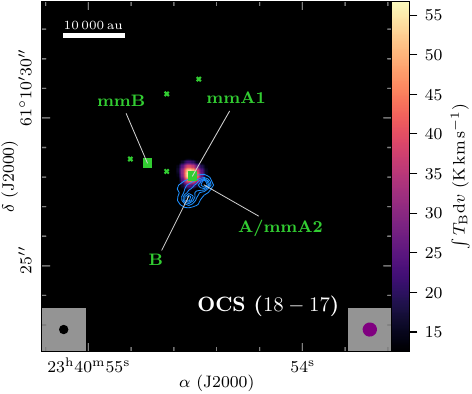}
\includegraphics[width=0.33\textwidth]{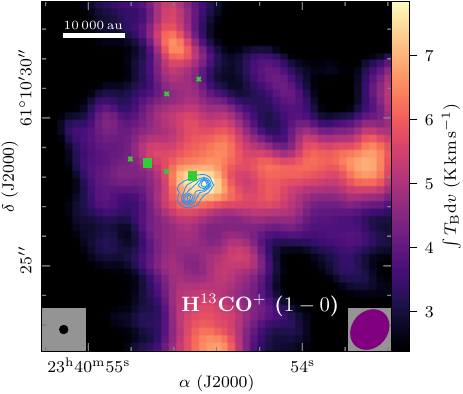}
\includegraphics[width=0.33\textwidth]{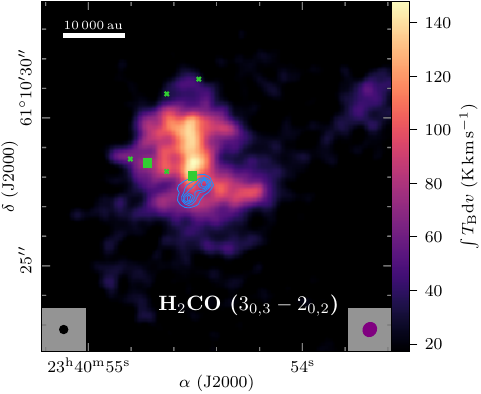}
\includegraphics[width=0.33\textwidth]{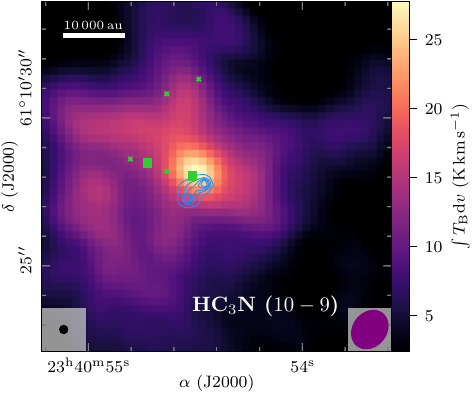}
\includegraphics[width=0.33\textwidth]{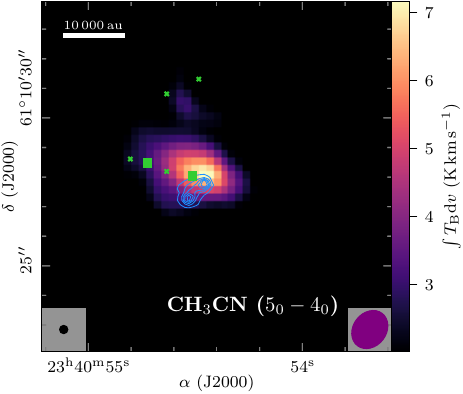}
\includegraphics[width=0.33\textwidth]{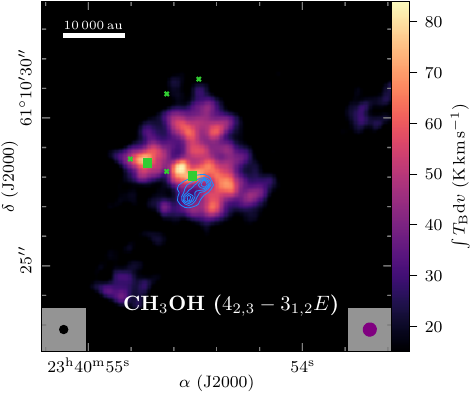}
\caption{Integrated intensity maps molecular lines detected with NOEMA plotted over the same area as the JWST data in Figs. \ref{fig:continuum} and \ref{fig:line}. In color, the line integrated intensity is shown. The JWST/MIRI 5.2\,$\upmu$m continuum is presented as blue contours with contour levels at 5, 10, 15, 20, 25$\times \sigma_{\mathrm{cont,5}\upmu\mathrm{m}}$. The two mm sources are indicated by green squares and several shock positions are highlighted by green crosses. In the bottom left and right corners, the synthesized beam of the NOEMA continuum and line data, respectively, is shown. In the top left panel, the bipolar outflows are indicated by red and blue dashed arrows \citep{Beuther2023}. The shock locations (Sect. \ref{sec:line}) and continuum sources are labeled in the top central and center left panel, respectively.}
\label{fig:line_mm}
\end{figure*}
	The integrated intensity maps of the mm lines are shown in Fig. \ref{fig:line_mm}. The bipolar outflows can partially also be revealed in the mm line data. A common shock tracer in the mm regime is SiO, with silicon (Si) being sputtered off the grains and forming SiO in the gas-phase \citep[e.g.,][]{Schilke1997}. The SiO emission peak is toward the northern part close to the shocked positions S2 and S3 also bright in H$_{2}$ and $[$S\,{\sc i}$]$. $^{13}$CS also shows an elongated structure toward the northern part. SO, H$^{13}$CO$^{+}$, and H$_{2}$CO show bright emission toward mmA1, but is also enhanced toward the northern direction toward S2 and S3. The north-south outflow (II) thus has a rich molecular component.
	
	In SiO, a large-scale outflow (I) is seen from the northwest to the southeast, which is also revealed by the CORE 1\,mm data \citep{Cesaroni2019}. This large-scale northwest-southeast molecular outflow (I) is also faintly seen in SO, H$_{2}$CO, and CH$_{3}$OH and is also detected in H$_{2}$ by JWST/MIRI (Fig. \ref{fig:continuum}). The S-shaped morphology of the outflow, most evident in SiO, suggests a precessing jet due to binary interaction \citep[e.g.,][]{Eisloffel1996,Fendt1998,Brogan2009,Sheikhnezami2015}.

	OCS shows very compact emission toward mmA1. HC$_{3}$N shows extended emission with many filamentary structures, hinting at not only outflowing but most likely also inflowing gas, and peaks toward mmA1. CH$_{3}$CN emission is strong around mmA1 and A/mmA2. Interestingly, CH$_{3}$OH emission significantly drops toward the locations of the MIR-bright sources A/mmA2 and B.
	
	In none of the NOEMA emission maps, mmB shows significant emission peaks, except for CH$_{3}$OH emission. Since all remaining detected continuum sources (mmA1, A/mmA2, and B) show protostellar activity through outflows and ionized gas, mmB is likely to be less evolved or less massive, although it cannot be completely ruled out that the H$_{2}$ and $[$S\,{\sc i}$]$ emission toward S1 could be an outflow originating from mmB that is directed along the line-of-sight.
	
	Shocks near the protostars are caused by the emerging outflows and are seen in SiO emission, but also in sulfur-bearing species, such as atomic sulfur, SO, and $^{13}$CS. However, OCS only emits strongly toward the source mmA1 suggesting that it may be related to ice sublimation, instead of shock chemistry. For SO, both high-temperature gas-phase chemistry close to the protostar and shock chemistry in the outflow (II) seem to be relevant to explain the spatial distribution. The entire region is embedded in a filamentary structure, seen in HC$_{3}$N, where most of these converge toward the central dense core. This suggests that even though there exist at least three outflows seen in H$_{2}$ emission by MIRI (I, II, and III, Fig. \ref{fig:continuum}), they have so far only little impact on disrupting the envelope in which the protostars are embedded in. Thus, further star formation activity could be still possible in the central dense core of IRAS\,23385 toward the dense millimeter source mmB that currently does not show any signatures of active star formation.

\subsection{Molecular hydrogen excitation diagram analysis}\label{sec:H2_modeling}

	The JWST/MIRI MRS spectral range covers multiple transitions of the pure rotational 0-0 state of H$_{2}$ (Table \ref{tab:MIRI_line_obs}) and with the excitation diagram analysis, the H$_{2}$ column density $N$(H$_{2}$) and rotational temperature $T_\mathrm{rot}$ can be estimated \citep[e.g.,][]{Neufeld2009}. The lowest H$_{2}$ transitions are excited through collisions \citep[e.g.,][]{Black1987}, while higher rotational levels and vibrational states can be excited through UV pumping. Thus the warm component linked to the collisions is a realistic tracer of the underlying gas temperature, while for any hotter component, the estimated rotation temperature may not correspond to a physical property. Since the angular resolution varies from 0\farcs2 to 0\farcs8 for 5\,$\upmu$m to 28\,$\upmu$m along the full MIRI spectral range, we smooth the S(7) to S(1) transitions to a common resolution of 0\farcs7 that corresponds to the angular resolution of the S(1) transition and regrid the remaining lines to the same spatial grid as the H$_{2}$ S(1) data. Spectral binning is not performed. This allows us to carry out a pixel-by-pixel excitation diagram analysis and to obtain a complete $N$(H$_{2}$) and $T_\mathrm{rot}$ map, since the H$_{2}$ transitions are detected within the entire FoV. 
	
	To provide reliable results, it is crucial to correct for extinction along the line-of-sight. We use an extinction curve derived by \citet{McClure2009} that is valid between a $K$-band (2.2\,$\upmu$m) magnitude $A_K$ of 1\,mag and 7\,mag. Since the extinction toward the central dense core is expected to be high \citep{Molinari1998B}, we assume in the entire field-of-view a $K$-band extinction of $A_K = 7$\,mag that roughly corresponds to a visual extinction of $\approx$50\,mag. We also tested lower values of $A_K$, but found that the integrated intensity of the S(3) transition that is located within the broad 10\,$\upmu$m silicon-absorption feature \citep{Beuther2023}, is significantly decreased compared to the transitions outside of the absorption feature for these lower values of $A_K$.
	
	As a first step, in each spatial pixel the seven H$_{2}$ transitions using the extinction-corrected data are fitted by a 1-dimensional (1D) Gaussian and from the integral of the Gaussian fit the integrated intensity of each transition is computed. Example spectra and Gaussian fits of the seven transitions taken from the position of source mmA1 and source B are presented in Fig. \ref{fig:h2_fit_example}. The integrated line intensities toward all four continuum sources (Table \ref{tab:sources}) and four shock positions (S1, S2, S3, and S4) are summarized in Table \ref{tab:H2_line_flux}.
	
	In general, the lines are barely resolved at the spectral resolution of MIRI. The uncertainty in the absolute flux calibration of the MIRI MRS instrument is only on the order of a few percent \citep{Argyriou2023}. A much higher uncertainty in the calculation of the line integrated intensities arises from the extinction-correction: We assume $A_K$ = 7 mag ($A_V$ = 54 mag) in the entire FoV, while there may be spatial differences. For comparison, we also present in Table \ref{tab:H2_line_flux} the extinction-corrected integrated line intensities calculated with $A_K$ = 5 mag ($A_V$ = 39 mag) and $A_K$ = 3 mag ($A_V$ = 23 mag). From the 9.7\,$\upmu$m silicate absorption feature $A_V = 30-40$ mag is estimated (Rocha et al., in prep.), whereas previous studies suggested even higher extinction values of $A_V$ = 200 mag \citep{Molinari1998B}. Given the high column densities toward the region, $A_K$ = $5-7$ mag should be a reliable estimate of the extinction. The line integrated line intensities vary between a factor of $2-3$ within this $K$-band extinction range (Table \ref{tab:H2_line_flux}). We thus assume in the following that the observed line integrated intensities are uncertain by a factor of two.
	
	Moreover, only transitions with a peak intensity of 50\,MJy\,sr$^{-1}$ (in the non-extinction corrected data) and above are further considered which is above the noise (Table \ref{tab:MIRI_line_obs}). With this threshold, we exclude areas with faint H$_{2}$ emission in the higher excited lines where the excitation diagram analysis would provide unreliable results. As an additional constraint, a fit to the excitation diagram is only performed when 5 or more transitions are detected in a spatial pixel above the threshold, as otherwise a two component fit would not provide reliable results.

\begin{figure}[!htb]
\centering
\includegraphics[width=0.49\textwidth]{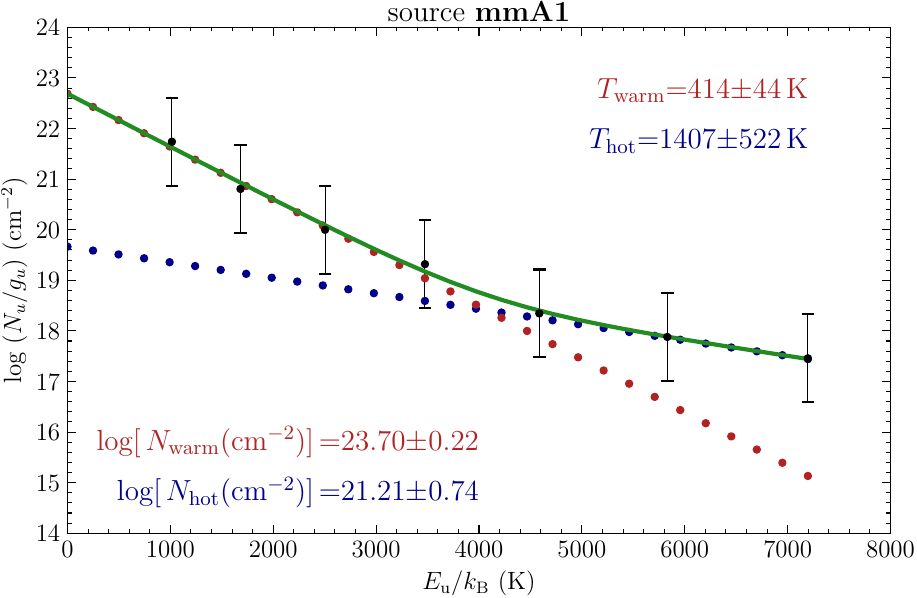}
\includegraphics[width=0.49\textwidth]{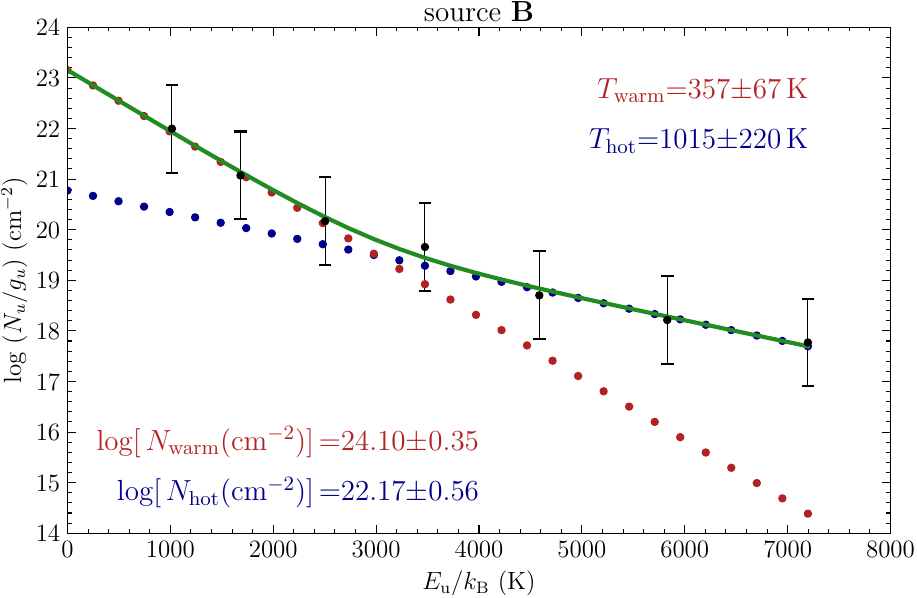}
\caption{Example of the H$_{2}$ excitation diagram analysis with \texttt{pdrtpy} of source mmA1 (top) and source B (bottom). The observed data is shown in black and the two component fit is shown by red and blue dots, corresponding to the warm and hot component, respectively and the total fit is indicated by a green line.}
\label{fig:h2_rot_dia}
\end{figure}
	
	The excitation diagram analysis is performed using the \texttt{pdrtpy}\footnote{\texttt{pdrtpy} is developed by Marc Pound and Mark Wolfire. This project is supported by NASA Astrophysics Data Analysis Program grant 80NSSC19K0573; from JWST-ERS program ID 1288 provided through grants from the STScI under NASA contract NAS5-03127; and from the SOFIA C+ Legacy Project through a grant from NASA through award \#208378 issued by USRA.} python package. Typically it is found that the observed H$_{2}$ transitions are best described by two rotational temperature components, which we refer to as $T_\mathrm{warm}$ and $T_\mathrm{hot}$ in the following. The observed line integrated intensities are converted into upper state column densities normalized by its statistical weight, $N_\mathrm{u}$/$g_\mathrm{u}$. The H$_{2}$ rotation temperature and column density can be estimated by plotting the logarithm of $N_\mathrm{u}$/$g_\mathrm{u}$, as a function of upper state energy level, $E_\mathrm{u}$/$k_\mathrm{B}$ and fitting the observed data points with a straight line for each component. The column density, log $N$, is proportional to the y-intercept and the slope is proportional to $-1/T$. We assume that all H$_{2}$ transitions are optically thin \citep[which is valid even for $N$(H$_{2}$)$>$10$^{23}$\,cm$^{-2}$,][]{Bitner2008}. 
	
	Examples of the excitation diagram toward the sources mmA1 and B are shown in Fig. \ref{fig:h2_rot_dia}. The fit results obtained from \texttt{pdrtpy} for the warm and hot component toward all four continuum sources (Table \ref{tab:sources}) and four shock positions (S1, S2, S3, and S4) are summarized in Table \ref{tab:H2_fit_results}. Considering all positions in the map, the uncertainties for the estimated temperature are between $10\%-30\%$, while for the column densities the uncertainties are on the order of a factor of $0.5-3$. In absolute numbers, the median uncertainties are log $\Delta N_\mathrm{warm} = 0.3$\,cm$^{-2}$, log $\Delta N_\mathrm{hot} = 0.6$\,cm$^{-2}$, $\Delta T_\mathrm{warm} = 43$\,K, $\Delta T_\mathrm{hot} = 270$\,K.
	
	The spectral line extraction and excitation diagram analysis toward the mmA1 and B sources shown in Figs. \ref{fig:h2_fit_example} and \ref{fig:h2_rot_dia}, respectively, reveal bright line emission in all transitions of H$_{2}$. The excitation diagram shows the necessity to fit the data with two temperature components. The warm and hot temperature component toward mmA1 are $\approx$400\,K and $\approx$1\,400\,K, respectively. The total column density, considering the contribution from both temperature components is $N_\mathrm{warm+hot} \approx 5\times10^{23}$\,cm$^{-2}$. Toward source B we find a higher column density, but lower temperature. This could be attributed to the fact that the lower excited H$_{2}$ lines may become optically thick toward the MIR sources, further discussed in Sect. \ref{sec:discussion}.
		
	In Table \ref{tab:H2_fit_results}, we also present excitation diagram fitting results with $A_K = 5$ mag and $A_K = 3$ mag assumed in the calculation of the extinction-corrected line integrated intensities (Table \ref{tab:H2_line_flux}). As expected, the estimated column densities are higher when the extinction is higher. However, given the high uncertainties, the differences are not significant. There is an effect on the derived temperatures. While the temperature of the hot component does not significantly change, the temperature of the warm component does show some differences, however not that significant given the high uncertainties. This is due to the fact that the S(3) line of H$_{2}$ lies in the silicate absorption feature and thus suffers from a higher extinction $A_\lambda$ compared to the remaining lines. Since this line lies at the border of the two components in the excitation diagram (Fig. \ref{fig:h2_rot_dia}), this data point does influence the fitted slope and thus the estimated temperature.
	
	The complete temperature and column density maps of the warm and hot components are presented in Fig. \ref{fig:Tmaps} and in Sect. \ref{sec:discussion} we compare the results with rotation temperature estimates based on the cold gas traced by the NOEMA data. One caveat mentioned before is that the excitation diagram analysis assumes that the H$_{2}$ lines are optically thin, which may especially not be the case toward the positions MIR sources A/mmA2 and B.

\subsection{Methyl cyanide line modeling}\label{sec:CH3CN_modeling}

	Molecular emission at mm wavelengths can be used to probe the temperature of the cold molecular gas. In \citet{Gieser2021}, the temperature of the cold gas component in the CORE regions, including IRAS\,23385, is estimated using H$_{2}$CO and CH$_{3}$CN emission lines. Unfortunately, the H$_{2}$CO lines are optically thick toward the center of IRAS\,23385 causing \texttt{XCLASS} to converge to unrealistic high temperatures $> 200$\,K. However, this highlights again that toward the central core, the densities and extinction must be high. For CH$_{3}$CN, the temperature toward mmA1 is estimated to be 100\,K \citep{Cesaroni2019,Gieser2021}, however, the transitions of the $J=12-11$ $K$-ladder at 1\,mm have high upper state energies and thus, the emission is compact around mmA1 and barely resolved. With the complementary 3\,mm CORE+ observations of the CH$_{3}$CN $J=5-4$ $K$-ladder, the temperature of more extended colder gas can be investigated (Table \ref{tab:mm_line_obs}).
	
	We model the rotational transitions of the CH$_{3}$CN $J=5-4$ $K$-ladder with \texttt{XCLASS}. With \texttt{XCLASS}, the 1D radiative transfer equation is solved assuming local thermal equilibrium (LTE) conditions. The fit parameters are source size $\theta_\mathrm{source}$, rotation temperature $T_\mathrm{rot}$, column density $N$(CH$_{3}$CN), line width $\Delta \varv$, and velocity offset $\varv_\mathrm{off}$. In total, we model four transitions of CH$_{3}$CN ($J=5-4$ and $K=0,1,2,3$, Table \ref{tab:mm_line_obs}) and from the best-fit, the rotation temperature can be estimated. Since these CH$_{3}$CN transitions have upper state energy levels $<$100\,K compared to the JWST/MIRI H$_{2}$ transitions with $>$1\,000\,K, we refer to this temperature component as the cold component, $T_\mathrm{cold}$. In each spatial pixel, the CH$_{3}$CN spectrum is modeled with \texttt{XCLASS} if the peak brightness temperature is higher than a threshold of 0.7\,K (corresponding to $\approx$3-4$\sigma_\mathrm{line}$, Table \ref{tab:mm_line_obs}).
	
\begin{figure*}[!htb]
\centering
\includegraphics[width=0.34\textwidth]{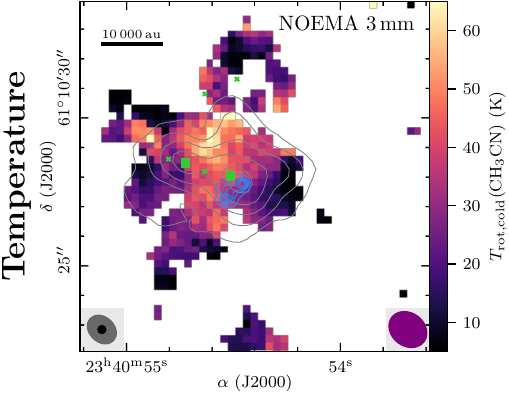}
\includegraphics[width=0.32\textwidth]{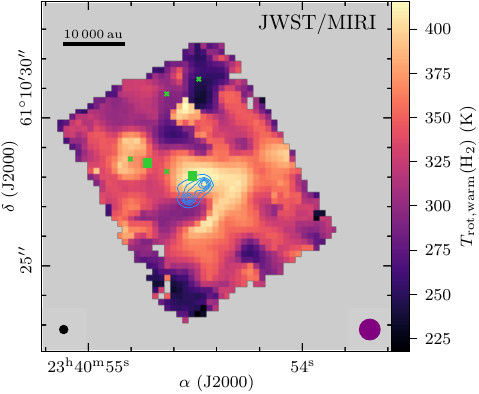}
\includegraphics[width=0.32\textwidth]{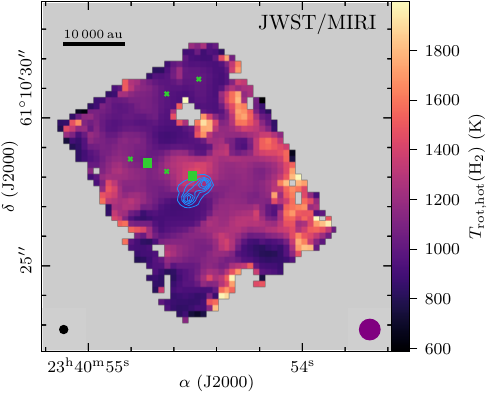}\\
\includegraphics[width=0.34\textwidth]{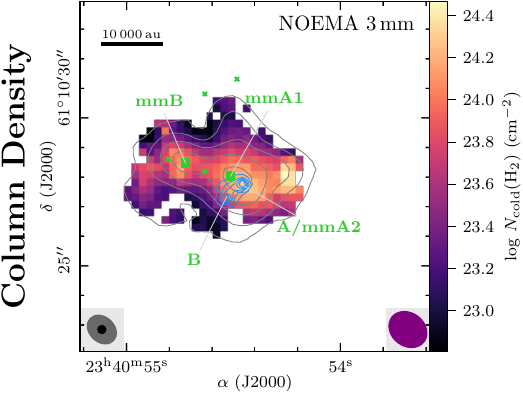}
\includegraphics[width=0.32\textwidth]{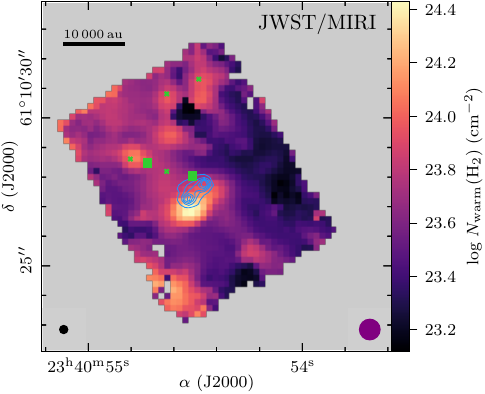}
\includegraphics[width=0.32\textwidth]{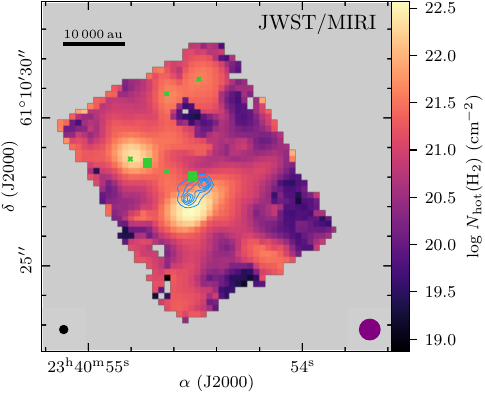}
\caption{Temperature and H$_{2}$ column density of the gas components in IRAS\,23385 derived using CH$_{3}$CN and H$_{2}$ as a diagnostic tool (Sects. \ref{sec:H2_modeling} and \ref{sec:CH3CN_modeling}). In the top and bottom panels, the rotation temperature and H$_{2}$ column density map, respectively, of the cold (\textit{left}), warm (\textit{center}), and hot (\textit{right}) components are shown in color. The angular resolution of the line data is indicated by a purple ellipse in the bottom right corner. The JWST/MIRI 5.2\,$\upmu$m continuum is presented by blue contours with contour levels at 5, 10, 15, 20, 25$\times \sigma_{\mathrm{cont,5}\upmu\mathrm{m}}$ and the angular resolution is highlighted by a black ellipse in the bottom left corner. The NOEMA 3\,mm continuum (top and bottom left panels) is highlighted by grey contours with levels at 5, 10, 20, 40, 80$\times \sigma_\mathrm{cont,3mm}$ and the synthesized beam is highlighted by a grey ellipse in the bottom left corner. All continuum sources are labeled in green in the bottom left panel and the mm continuum sources are marked by green squares. Several shock positions are indicated by green crosses (Sect. \ref{sec:line}).}
\label{fig:Tmaps}
\end{figure*}

	As an example, the observed spectrum and best-fit model derived with \texttt{XCLASS} is shown in Fig. \ref{fig:ch3cn_fit_example} toward the position of mmA1. The derived rotation temperature, column density, and line width of CH$_{3}$CN is 46\,K, $6.4\times10^{13}$\,cm$^{-2}$, and 3.7\,km\,s$^{-1}$, respectively. The full temperature map is presented in Fig. \ref{fig:Tmaps} and is discussed and compared to the temperature estimated from H$_{2}$ in Sect. \ref{sec:discussion}. 
	
	Using the CH$_{3}$CN temperature and the 3\,mm continuum data, we can also estimate the H$_{2}$ column density of the cold component, if the 3\,mm continuum stems from optically thin dust emission and if the gas and dust temperatures are coupled. The 3\,mm continuum optical depth can be calculated according to
	\begin{equation}
	\tau_\mathrm{cont,3mm} = -\mathrm{ln} \left(1-\frac{I_\mathrm{pix,3mm}}{\Omega B_\nu (T_\mathrm{cold})}\right).
	\end{equation}
	$I_\mathrm{pix,3mm}$ is the 3\,mm intensity in a pixel, converted from Jy\,beam$^{-1}$ to Jy\,pixel$^{-1}$ considering the beam area $\Omega = \pi/(4\mathrm{ln}2) \theta_\mathrm{maj} \theta_\mathrm{min}$ and the area covered by one pixel (0.25$''$)$^2$, and $B_\nu(T_\mathrm{cold})$ is the Planck function. We find that $\tau_\mathrm{cont,3mm} < 0.02$ in the entire map. Thus, the 3\,mm continuum emission stems from optically thin dust emission.
	
	Using the 3\,mm continuum data, that has a similar angular resolution, and the CH$_{3}$CN temperature map, the H$_{2}$ column density of the cold gas can be derived \citep{Hildebrand1983}:
	
	\begin{equation}
	\label{eq:N}
	N_\mathrm{cold}(\mathrm{H}_{2}) = \frac{I_\mathrm{pix,3mm} \gamma}{B_\nu(T_\mathrm{cold}) \Omega \kappa_\nu \mu m_\mathrm{H}}.
	\end{equation}
	
	We assume a gas-to-dust mass ratio of $\gamma = 150$ \citep[see][]{Beuther2018}, a dust opacity of $\kappa_\nu = 0.17$\,cm$^{2}$\,g$^{-1}$ \citep[Table 1 in][extrapolated from 1.3\,mm, and assuming thin icy mantles at a gas density of $10^6$\,cm$^{-3}$]{Ossenkopf1994}, a mean molecular weight of $\mu = 2.8$ and $m_\mathrm{H}$ is the mass of a hydrogen atom. The flux calibration of NOEMA is expected to be accurate to a 10\% level at 3\,mm\footnote{\url{https://www.iram.fr/IRAMFR/GILDAS/doc/html/pdbi-cookbook-html/pdbi-cookbook.html}}. The temperatures estimated with CH$_{3}$CN in \texttt{XCLASS} have uncertainties on the order of 30\% \citep{Gieser2021}. The biggest uncertainty when estimating $N_\mathrm{cold}$ lies in the gas-to-dust mass ratio $\gamma$ and the dust opacity $\kappa_\nu$. It can thus be estimated that the calculated $N_\mathrm{cold}$ values can be uncertain up to a factor of $2-4$ \citep{Beuther2018}. The results are for $T_\mathrm{cold}$, $M_\mathrm{cold}$, and $N_\mathrm{cold}$(H$_{2}$) are presented in Fig. \ref{fig:Tmaps} and are discussed in Sect. \ref{sec:discussion}.
	
\section{Discussion}\label{sec:discussion}

	The JWST/MIRI MRS observations of IRAS\,23385 reveal exciting new insights into energetic processes during the formation of stars. Multiple outflows (I, II, and III, Fig. \ref{fig:continuum}) can be identified, one of which shows in addition to H$_{2}$ and $[$S\,{\sc i}$]$ emission, enhanced emission by $[$Fe\,{\sc ii}$]$, $[$Ni\,{\sc ii}$]$, and $[$Ne\,{\sc ii}$]$. The outflows are also associated with cold molecular gas traced by mm lines. Here, we focus on a few main points based on the analysis (Sect. \ref{sec:results}) and discuss in detail the different gas components, the outflow properties, and the spatial emission and intensity ratio of $[$Ne\,{\sc ii}$]$ and $[$Ne\,{\sc iii}$]$.

\subsection{Gas components toward the central dense core}

	A comparison of the warm and hot gas traced by H$_{2}$ in the JWST/MIRI observations (Sect. \ref{sec:H2_modeling}) and the cold gas traced by CH$_{3}$CN and the 3\,mm continuum in the NOEMA observations (Sect. \ref{sec:CH3CN_modeling}) is presented in Fig. \ref{fig:Tmaps}. The top and bottom panels show the H$_{2}$ rotation temperature and column density, respectively, of the cold (left), warm (center), and hot component (right). To our knowledge, this is the first time that spatially resolved maps of the cold, warm and hot components have been obtained in a high-mass cluster-forming region. The cold component consists of dense (10$^{23}$-10$^{24}$\,cm$^{-2}$) and cold (10-60\,K) gas. Interestingly, for the warm component we derive similar column densities, but spatially different, since the H$_{2}$ lines trace shocked and heated gas due to the outflows, while the mm line and continuum traces the cold envelope. The temperature of the warm component ranges between 200-400\,K. 
	
	Comparing the temperature and column density, we find that these properties are anti-correlated, such that toward the high column densities the temperature is lower compared to the lower column density regions. This hints that toward the dense shocks, the lower H$_{2}$ lines could become optically thick and we only trace outer layers toward these shock positions. In the hot component, the column densities are about two orders of magnitude lower, $N_\mathrm{hot}\approx 10^{22}$\,cm$^{-2}$, compared to the cold and warm component and the temperatures are 1\,000-1\,800\,K.

	How do these temperatures and column densities compare with those found in nearby low-mass regions for which more extensive studies are available in the literature? Taking as an example the L1157 Class 0 protostar \citep[$L = 8.4$\,$L_\odot$,][]{Froebrich2005}, \citet{DiFrancesco2020} studied the cold material using \textit{Herschel} observations and find dust temperatures of 10-18\,K and column densities up to $2\times10^{22}$\,cm$^{-2}$. \citet{Nisini2010} created temperature and H$_{2}$ column density maps of the L1157 outflow and also find that the H$_{2}$ line fluxes are best described by two temperature components, with $300-400$\,K and $1\,100-1\,400$\,K, respectively. These results are similar to the temperatures derived in the IRAS\,23385 region (Fig. \ref{fig:Tmaps}), however, the H$_{2}$ column density map derived by \citet{Nisini2010} is about four orders of magnitude lower as expected in less massive outflows driven by low-mass protostars. Compared to previous IR studies toward HMSFRs, with JWST we are for the first time able to probe high-column density regions ($>10^{23}$\,cm$^{-2}$) due to the higher angular resolution. The temperature in IRAS\,23385 is also similar to low-mass protostars in NGC 1333 \citep{Dionatos2020}. In the regime of high-mass protostars, \citet{vandenAncker2000} find temperatures of the warm component of 500 \,K and 730\,K for the HMSFR S106 and Cep A, respectively. In the Orion OMC-1 outflow, \citet{Rosenthal2000} find that bulk ($\approx$ 72\%) of warm gas has a temperature of $\approx$630\,K. 
	
	A system that is comparable to IRAS\,23385 is IRAS\,20126$+$4104 \citep[$d = 1.7$\,kpc,][]{Cesaroni1997} with a similar mass of the central protostar (7\,$M_\odot$) and luminosity ($10^4\,L_\odot$). Jet precession in this source also hints at underlying multiplicity \citep[e.g.,][]{Cesaroni2005, CarattioGaratti2008}. Using higher excited transitions of H$_{2}$ in the NIR a high temperature component at 2\,000\,K is derived by these authors, while also a warm component is found at 500\,K. Additional observations in the NIR, for example with the Near-Infrared Spectrograph (NIRSpec) instrument of JWST, are required in order to investigate the presence of potentially even higher temperature components in IRAS\,23385.
	
	Since toward IRAS\,23385 the highest mass protostar is estimated to be 9\,$M_\odot$ \citep{Cesaroni2019}, at the lower end of what is considered as a ``high-mass'' star, it is expected that more massive protostars can heat up the environment to even higher temperatures through powerful outflows and expanding UCH{\sc ii} regions. The fact that a cold component as traced by the NOEMA mm observations still exists toward the central dense core, stresses that star formation can efficiently occur in shielded dense gas. While outflows seem to impact and surrounding envelope through shocks and heating of the material, the feedback is not (yet) strong enough to completely dissipate the central dense core as outflows are collimated.
	 
\subsection{Composition of the outflows and evolutionary stage of the continuum sources}

\begin{figure*}[!htb]
\includegraphics[width=0.99\textwidth]{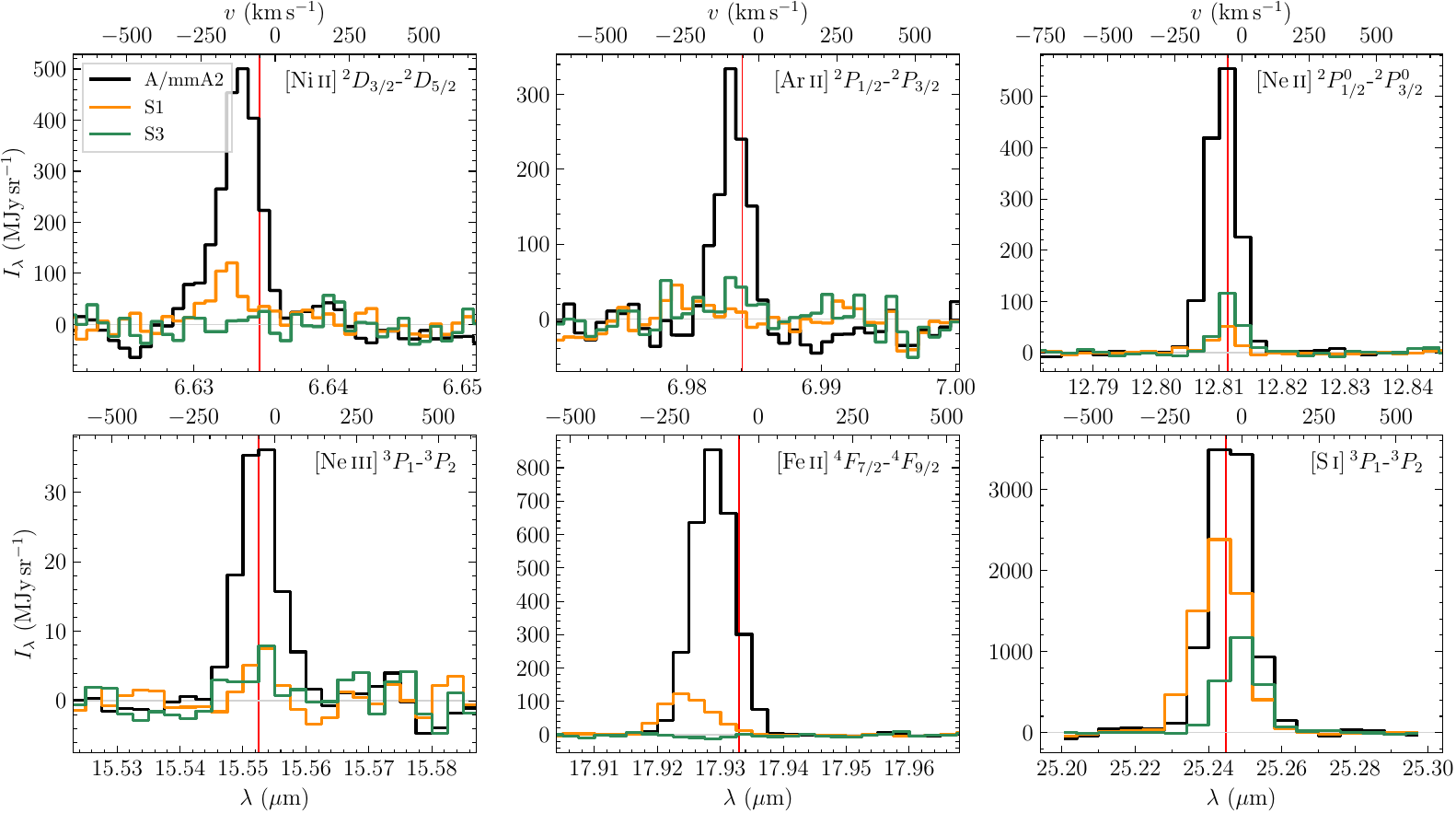}
\caption{JWST/MIRI MRS spectra of atomic line emission. In each panel, the spectra are presented extracted from the positions at the MIR source A/mmA2 (black, Table \ref{tab:sources}) and two shock locations (orange: S1 and green: S3, see Fig. \ref{fig:line}). The bottom and top x-axis show the observed wavelength and velocity, respectively. The red vertical line indicates the wavelength of the line at a source velocity of $-50.2$\,km\,s$^{-1}$.}
\label{fig:spectra}
\end{figure*}

	The northeast-southwest outflow III (Fig. \ref{fig:continuum}) is not only seen in H$_{2}$ and $[$S\,{\sc i}$]$ emission knots, but these knots are also bright in $[$Fe\,{\sc ii}$]$ and $[$Ni\,{\sc ii}$]$ emission (Fig. \ref{fig:line}). \citet{Dionatos2014} find $[$Fe\,{\sc ii}$]$ and $[$Ne\,{\sc ii}$]$ to trace the FUV irradiation on the disk surface and on the walls of the outflow cavity and \citet{CarattioGaratti2015} also find that $[$Fe\,{\sc ii}$]$ and H$_{2}$ knots can be co-spatial. Toward the position of the source A/mmA2 we detect emission from ionized atoms from $[$Fe\,{\sc ii}$]$, $[$Ni\,{\sc ii}$]$, $[$Ne\,{\sc ii}$]$, $[$Ne\,{\sc iii}$]$, and $[$Ar\,{\sc ii}$]$ (Fig. \ref{fig:line}). This suggests that this protostar is the most evolved with strong UV and X-ray radiation. The ionization energies are 7.6\,eV (Ni), 7.9\,eV (Fe), 15.8\,eV (Ar), and 21.6\,eV (Ne). In Fig. \ref{fig:spectra}, example spectra of the atomic transitions extracted from the positions of the continuum source A/mmA2 and the shock positions S1 and S3 are shown. The $[$S\,{\sc i}$]$ line peaks at the source velocity toward source A/mmA2 and shock position S1. The shock position S3 caused by the redshifted part of outflow II is as expected redshifted. If detected, $[$Ni\,{\sc ii}$]$, $[$Ar\,{\sc ii}$]$, $[$Fe\,{\sc ii}$]$ lines are blueshifted compared to the source velocity tracing the blueshifted outflow components. The $[$Ne\,{\sc ii}$]$ and $[$Ne\,{\sc iii}$]$ lines are further discussed in Sect. \ref{sec:Nediscussion}.
	

	\citet{Rosenthal2000} studied the MIR spectrum of one of the brightest H$_{2}$ shock position of the Orion OMC-1 outflow. In comparison with the IRAS\,23385 outflows, these authors find emission from double or triple ionized fine structure lines by $[$Fe\,{\sc iii}$]$, $[$S\,{\sc iii}$]$, $[$P\,{\sc iii}$]$, $[$S\,{\sc iv}$]$, and $[$Ar\,{\sc iii}$]$, while in IRAS\,23385 only $[$Ne\,{\sc iii}$]$ is detected (the emission of this line is further discussed in the following). This suggests that the protostars and their outflows in IRAS\,23385 are not energetic enough to produce and excite the double or higher ionized transitions.
	
	We do not detect MIR continuum emission toward the mm continuum peak source mmA1. Either the extinction is too high toward the location of this source, for example, due to the geometry or high density, or the protostar itself is at an earlier evolutionary phase compared to the MIR-bright sources A/mmA2 and B. However, we could expect this source to become bright at the longer wavelengths within the MIRI coverage, which is not the case. The bipolar outflow (II) in the north-south direction only consists of molecules and neutral atomic lines (see Fig. \ref{fig:spectra} for the atomic line emission at the shock position S1). The fact that CH$_{3}$CN and OCS emission is detected toward the location of mmA1, suggests that the central protostar is currently heating up the surrounding material creating a hot core region where ices sublimate from the grains. The high density may currently still prevent energetic UV and X-ray radiation and also the MIR radiation of the protostar itself to escape far from the protostar and thus remains undetected in the MIR.
	
	The source mmB does not show any signatures of star-forming activity. In fact, the CH$_{3}$OH emission detected toward the location of mmB may be shocked gas caused by the northeast-southwest outflow (III). Though it may not be a coincidence that mmB is along this outflow direction as it may be cold material that was swept up by the outflow (III).
	
	\citet{Tychoniec2021} created an overview of molecular lines that trace different physical entities surrounding low-mass Class 0 and I protostars, for example, the hot core region, outflows, UV-irradiated cavity walls, jets, disks, and the surrounding envelope (see their Fig. 11 and Table 2). All molecular lines at mm wavelengths presented in this work (Fig. \ref{fig:line_mm}), except for HC$_{3}$N, are covered by their overview. We find a similar spatial distribution in IRAS\,23385, though due to the large distance we are not able to resolve a potential disk or differentiate between the outflow and cavity walls. Interestingly, we do not find strong emission by complex organic molecules (COMs), except for CH$_{3}$OH and CH$_{3}$CN. The observed CH$_{3}$OH emission traces the outflow (I), it is non-thermally desorbed and likely sputtered as observed in \citet{Perotti2021}. Given that the temperature of the cold component is $\approx$60\,K, the COMs could still mostly reside in ices on the dust grains. The composition of the MIRI MRS ice absorption spectra could reveal the presence of COMs in the ices and will be presented in Rocha et al. (in prep.). Early JWST observations toward low-mass protostars and their envelopes have already revealed a variety of different ice species \citep{Yang2022,McClure2023}.
	
		For both outflows III and II (Fig. \ref{fig:continuum}), the atomic sulfur is co-spatial with H$_{2}$ knots caused by shocks due to the outflow (Fig. \ref{fig:line}). This suggests that sulfur is being released by $C$-type shocks \citep[similar to outflows of Class 0 protostars,][]{Anderson2013}. Since the observed $[$S\,{\sc i}$]$ line at 25.2\,$\upmu$m has a large critical density \citep[10$^{5}$\,cm$^{-3}$,][]{Hollenbach1989}, it will be readily detected in high-density shocked gas. Its abundance can be enhanced with sputtering of the grains in shocks.
		
	Shocks in star-forming regions can occur through various processes, e.g., cloud-cloud collisions, jets and molecular outflows, stellar winds, or expanding H{\sc ii} regions. In $C$-type shocks, the maximum temperature is 2\,000$-$3\,000\,K and shock velocities are $<$40\,km\,s$^{-1}$, that is too low to destroy molecules, while in $J$-type shocks temperatures can reach $10^{5}$\,K with increased fine structure emission from atoms such as Fe, Mg, and Si that sputtered off the grains and shock velocities that are typically higher than 40\,km\,s$^{-1}$\citep{Hollenbach1989,vanDishoeck2004}. With the relatively low spectral resolution of JWST/MIRI MRS ($R \approx 3\,700 - 1\,400$) it is difficult to estimate the shock velocities, however, the presence of $[$Fe\,{\sc ii}$]$ \citep[velocity-resolved maps from $-$185\,km\,s$^{-1}$ to 120\,km\,s$^{-1}$ are presented in][their Fig. 6]{Beuther2023} and $[$Ni\,{\sc ii}$]$ in the northeast-southwest outflow (III) suggests that there could be a contribution of $J$-type shocks, while for the remaining two outflows (II and I) only enhanced emission of H$_{2}$ and $[$S\,{\sc i}$]$ is detected, and ionized atomic fine structure lines are absent toward these outflow directions.

\subsection{PDRs vs. shocks: The case of $[$Ne\,{\sc ii}$]$ and $[$Ne\,{\sc iii}$]$}\label{sec:Nediscussion}

	Previous ISO observations have suggested that $[$Ne\,{\sc ii}$]$ could be tracing dissociative ($J$-type) shocks \citep{vanDishoeck2004}. With the low angular resolution of these older data often multiple physical entities are covered, for example, H{\sc ii} regions, photon-dominated/photon-dissociation regions (PDRs), and outflows \citep[e.g. in Orion IRc2][]{vanDishoeck1998}. In addition, due to low filling factors toward distant HMSFRs such lines were often not detected. With JWST/MIRI MRS we are able to resolve individual components. In fact, both shocks and PDRs can contribute to many of the lines observed in this work.
	
	The spatial morphology of the $[$Ne\,{\sc ii}$]$ line (Fig. \ref{fig:line}) reveals extended emission within the full MIRI FoV, and is also detected toward the east at the arc-like structure that is a PDR illuminated by the nearby UCH{\sc ii} region. In PDRs the radiation field is dominated by FUV photons (6-13.6\,eV) \citep[e.g.][]{Hollenbach1999}. PDRs show emission by MIR H$_{2}$ transitions, as well as neutral and single ionized atomic fine structure lines \citep[][and references within]{vanDishoeck2004}. In the case of the IRAS\,23385 PDR, we detect emission of H$_{2}$ and $[$Ne\,{\sc ii}$]$ in the arc highlighting the impact of the clustered mode of high-mass star-formation including photons with energies $>$13.6\,eV that can ionize Neon.
	
	The $[$Ne\,{\sc ii}$]$ transition also traces both outflows and the inner disk toward low-mass protostars where these regions can be sufficiently resolved \citep{Sacco2012}. These authors find that the $[$Ne\,{\sc ii}$]$ emission is slightly blueshifted (2$-$12\,km\,s$^{-1}$) and argue that it may be caused by a disk wind. \citet{Glassgold2007} suggest that the emission could stem from the disk that is irradiated by stellar X-rays. $[$Ne\,{\sc ii}$]$ is commonly detected in Class II disks \citep{Pascucci2022}. While most atomic fine structure lines are indeed blueshifted by 50$-$100\,km\,s$^{-1}$ with respect to the source velocity, the $[$Ne\,{\sc ii}$]$ and $[$Ne\,{\sc iii}$]$ lines toward the position of source A/mmA2 are not blueshifted but more centered on the source velocity (Fig. \ref{fig:spectra}), so the emission at source A/mmA2 could either stem from the protostar itself or a disk wind.
	
	Considering the $[$Ne\,{\sc ii}$]$ and $[$Ne\,{\sc iii}$]$ peak toward the west of source A/mmA2 (Fig. \ref{fig:line}), a disk wind is unlikely to explain the emission since it is $\approx$ 5\,000\,au offset from the MIR source. Another explanation could be the that we are tracing the UV and X-ray irradiated cavity of the large scale northwest-southeast outflow (I) launched from source B. However, the emission is not perfectly aligned with that direction. Since most emission lines toward source A/mmA2 shown in Fig. \ref{fig:spectra} are blueshifted compared to the source velocity, while $[$Ne\,{\sc ii}$]$ and $[$Ne\,{\sc iii}$]$ peak at the source velocity, this suggests that $[$Ne\,{\sc ii}$]$ and $[$Ne\,{\sc iii}$]$ are tracing a region closer to the protostar. It remains inconclusive what causes the $[$Ne\,{\sc ii}$]$, $[$Ne\,{\sc iii}$]$, and faint $[$Ar\,{\sc ii}$]$ emission to the west of source A/mmA2 (Fig. \ref{fig:line}).
	
	
	
	High-mass (proto)stars have enough UV-intensity to ionize $[$Ne\,{\sc ii}$]$ and $[$Ne\,{\sc iii}$]$. To estimate the line integrated flux density of both $[$Ne\,{\sc ii}$]$ and $[$Ne\,{\sc iii}$]$ transitions, we perform aperture photometry with an aperture of 0\farcs9 toward both $[$Ne\,{\sc iii}$]$ peak positions (highlighted by grey dash-dotted circles in Fig. \ref{fig:line}). Toward source A/mmA2 the line integrated flux density is 3.1\,mJy for $[$Ne\,{\sc ii}$]$ and 0.23\,mJy for $[$Ne\,{\sc iii}$]$, while to the western position it is 2.6\,mJy and 0.19\,mJy, respectively. While the flux density is higher at the position of source A/mmA2, the $[$Ne\,{\sc ii}$]$/$[$Ne\,{\sc iii}$]$ flux density ratio at both positions is similar (13 and 14) suggesting that even at a projected distance of 5\,000\,au, the X-ray and UV-radiation from the protostar can efficiently funnel along the outflow. \citet{Simpson2012} find $[$Ne\,{\sc ii}$]$/$[$Ne\,{\sc iii}$]$ ratios between 17 and 125 in a sample of massive protostars in the G333.2$-$0.4 giant molecular cloud and \citet{Rosenthal2000} find a ratio of 0.95 toward the brightest outflow knot in the Orion OMC-1 outflow. Toward the Orion IRc2 source that causes this outflow, a ratio of 1.4 is derived \citep{vanDishoeck1998}. In the Orion PDR a ratio of 1.3-1.9 is derived. \citet{Guedel2010} find a higher $[$Ne\,{\sc ii}$]$ luminosity in sources associated with jets compared to those without which would be consistent with our high ratio. These authors find a weak correlation with X-ray luminosity. When the full sample of the JOYS targets is observed, a statistical analysis of such properties can be conducted from low- to high-mass protostars and contributions for jets versus UV and/or X-rays can be disentangled.

\section{Conclusions}\label{sec:conclusions}

	In this study we compare the MIR properties of the warm gas in IRAS\,23385 traced by JWST/MIRI MRS observations that are part of the JOYS project with the cold material traced by NOEMA observations at mm wavelengths. In total, four continuum sources are resolved that are all embedded within the dense cold envelope. Two sources are MIR bright, while the two bright mm cores remain undetected in the MIRI range between 5 and 28\,$\upmu$m. The spatial morphology of the atomic and molecular lines is investigated using integrated intensity maps. The gas temperature and column density of different components are estimated using H$_{2}$ MIR and CH$_{3}$CN mm line emission and 3\,mm continuum emission. Our conclusions are summarized as follows:

	\begin{enumerate}
		\item Combining the JWST/MIRI MRS and NOEMA observations allows us to disentangle several gas components, consisting of cold ($<$100\,K), warm (200-400\,K) and hot ($\geq$1\,000\,K) material, traced by CH$_{3}$CN and H$_{2}$ lines. While the warm and hot component, mostly caused by heated and/or shocked gas in the outflows, is extended, there is still a significant amount of cold material that allows for star formation to proceed. This is also highlighted by the H$_{2}$ column densities of the cold and warm components that are both on the order of 10$^{24}$\,cm$^{-2}$. The column density of the hot component is in contrast significantly lower, on the order of 10$^{22}$\,cm$^{-2}$.
		\item The H$_{2}$ 0-0 S(1) to S(7) transitions reveal $C$-type shocked and extended warm gas due to multiple bipolar outflows (I, II, and III). These are also evident in the atomic sulfur line $[$S\,{\sc i}$]$, stressing the importance of sulfur being released from the grains due to shocks. The outflows and their cavities can also be seen in many molecular lines at mm wavelengths (SiO, SO, $^{13}$CS, H$^{13}$CO$^{+}$, H$_{2}$CO, CH$_{3}$OH).
		\item An energetic outflow (III) emerges probably from source A/mmA2 where the blueshifted outflow reveals emission knots by $[$Fe\,{\sc ii}$]$ and $[$Ni\,{\sc ii}$]$. Thus, the outflow seems to have undergone multiple outbursts in the past causing dissociative $J$-type shocks. 
		\item The MIR source A/mmA2 also shows enhanced emission toward the central location of the protostar by $[$Fe\,{\sc ii}$]$, $[$Ni\,{\sc ii}$]$, as well as $[$Ne\,{\sc ii}$]$, $[$Ne\,{\sc iii}$]$, and $[$Ar\,{\sc ii}$]$. While the emission is blueshifted by 50$-$100\,km\,s$^{-1}$ compared to the source velocity for most of these species, the emission from $[$Ne\,{\sc ii}$]$, and $[$Ne\,{\sc iii}$]$ peaks at the source velocity. This suggests that in these cases, not only the outflow may cause the ionization and excitation of the line emission, but, for example, a disk wind or the disk atmosphere itself or an X-ray irradiated cavity, is the cause. However, the angular and spectral resolution are not sufficient to differentiate between these cases.
		\item The $[$Ne\,{\sc ii}$]$ emission is extended within the full MIRI FoV and also traces bright emission toward the edges of the MIRI FoV, a PDR region caused by the nearby UCH{\sc ii} regions (Fig. \ref{fig:overview}). 
	\end{enumerate}
	
	To fully understand the clustered properties of high-mass star formation it is important to conduct a multi-wavelength study tracing both the warm and the cold gas properties. With JWST we now have a new unprecedented view on the IR properties and diagnostics of protostars and their environments at high angular resolution and sensitivity. 

\begin{acknowledgements}
	This work is based on observations made with the NASA/ESA/CSA James Webb Space Telescope. The data were obtained from the Mikulski Archive for Space Telescopes at the Space Telescope Science Institute, which is operated by the Association of Universities for Research in Astronomy, Inc., under NASA contract NAS 5-03127 for JWST. These observations are associated with program \#1290. The following National and International Funding Agencies funded and supported the MIRI development: NASA; ESA; Belgian Science Policy Office (BELSPO); Centre Nationale d’Etudes Spatiales (CNES); Danish National Space Centre; Deutsches Zentrum fur Luftund Raumfahrt (DLR); Enterprise Ireland; Ministerio De Economi\'a y Competividad; Netherlands Research School for Astronomy (NOVA); Netherlands Organisation for Scientific Research (NWO); Science and Technology Facilities Council; Swiss Space Office; Swedish National Space Agency; and UK Space Agency. This work is based on observations carried out under project number L14AB and W20AV with the IRAM NOEMA Interferometer and the 30m telescope. IRAM is supported by INSU/CNRS (France), MPG (Germany) and IGN (Spain). H.B.~acknowledges support from the Deutsche Forschungsgemeinschaft in the Collaborative Research Center (SFB 881) “The Milky Way System” (subproject B1). EvD, MvG, LF, KS, WR and HL acknowledge support from ERC Advanced grant 101019751 MOLDISK, TOP-1 grant 614.001.751 from the Dutch Research Council (NWO), the Netherlands Research School for Astronomy (NOVA), the Danish National Research Foundation through the Center of Excellence “InterCat” (DNRF150), and DFG-grant 325594231, FOR 2634/2. P.J.K.~acknowledges financial support from the Science Foundation Ireland/Irish Research Council Pathway programme under Grant Number 21/PATH-S/9360. A.C.G. has been supported by PRIN-INAF MAIN-STREAM 2017 “Protoplanetary disks seen through the eyes of new-generation instruments” and from PRIN-INAF 2019 “Spectroscopically tracing the disk dispersal evolution (STRADE)”. K.J.~acknowledges the support from the Swedish National Space Agency (SNSA). T.H.~acknowledges support from the European Research Council under the Horizon 2020 Framework Program via the ERC Advanced Grant "Origins" 83 24 28. T.P.R. acknowledges support from ERC grant 743029 EASY.
\end{acknowledgements}

\bibliographystyle{aa}
\bibliography{IRAS23385_JOYS_paper2_draft_v9}

\begin{appendix}

\section{Fit examples and results}\label{app:fitexamples}

\begin{figure}[!htb]
\centering
\includegraphics[width=0.49\textwidth]{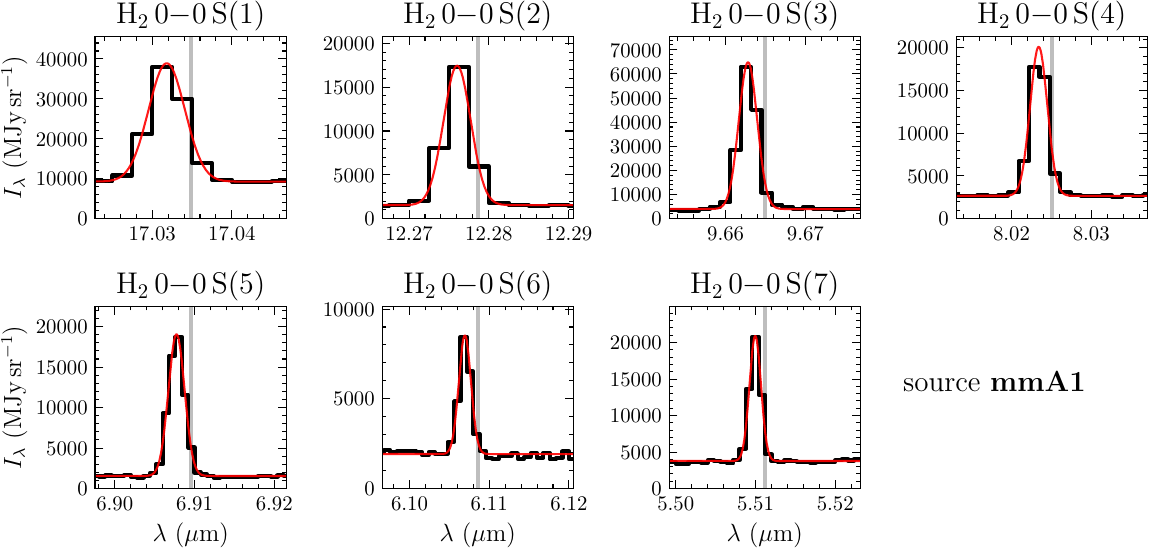}
\includegraphics[width=0.49\textwidth]{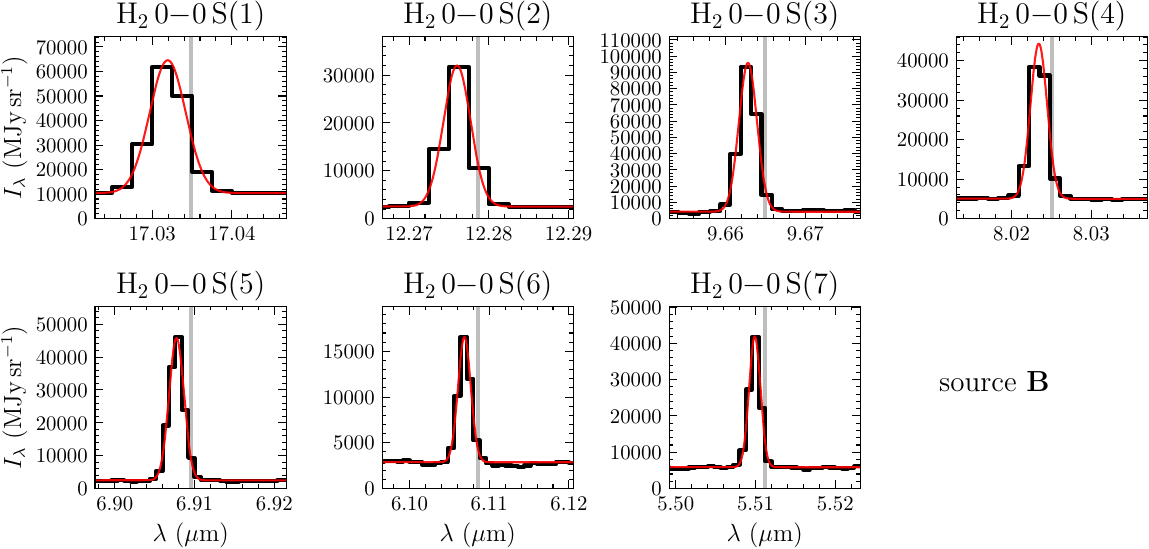}
\caption{Examples of the observed and extinction-corrected H$_{2}$ lines (black) and Gaussian fit (red line) of source mmA1 (top) and source B (bottom). The vertical grey line indicates the wavelength of each transition.}
\label{fig:h2_fit_example}
\end{figure}

	In Sect. \ref{sec:H2_modeling} the H$_{2}$ 0-0 S(1) to S(7) transitions covered by JWST/MIRI are used in order to estimate the temperature of the warmer gas. All lines are smoothed to a common angular resolution of 0\farcs7 and spatially regridded to the S(1) line data. In each spatial pixel, the spectra are extracted and fitted by a Gaussian + constant component to include the continuum emission. The line integrated intensities are listed in Table \ref{tab:H2_line_flux} for all continuum sources and shock positions assuming $A_K = 7$ mag in the calculation of the extinction corrected line integrated intensities. For comparison, we also show the results for $A_K = 5$ mag and $A_K = 3$ mag. Figure \ref{fig:h2_fit_example} shows the observed spectra and corresponding Gaussian fit toward the positions of source mmA1 (top panel) and source B (bottom panel). The spectra are corrected for extinction using the extinction curve derived by \citet{McClure2009}.
	
	The line integrated intensities of the Gaussian fits are converted to upper state column densities divided by the statistical weight and when plotted against the upper energy level, the H$_{2}$ rotation temperature and column density can be estimated from the slope and intercept, respectively. Examples of the H$_{2}$ excitation diagrams are shown in Fig. \ref{fig:h2_rot_dia} for sources mmA1 and B. Typically for star-forming regions, the profile is best described by two temperature components, which we refer to as the ``warm'' and ``hot'' component in contrast to the ``cold'' component traced by the NOEMA data. The fit results for the warm and hot component are summarized in Table \ref{tab:H2_fit_results} for all continuum sources and shock positions. For comparison, we also present the results using extinction values of $A_K = 5$ mag and $A_K = 3$ mag. There are no significant differences for the derived column densities. A small effect is seen for the temperatures given that the S(3) transition that lies in the silicate absorption feature influences the two fitted slopes.
	
\begin{figure}[!htb]
\centering
\includegraphics[width=0.45\textwidth]{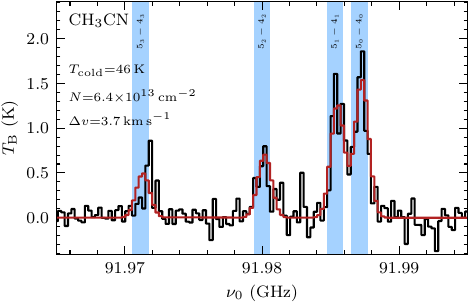}
\caption{Example of CH$_{3}$CN line modeling with \texttt{XCLASS}. In black, the observed spectrum toward mmA1 is shown and in red the best-fit model derived by \texttt{XCLASS}. The modeled transitions of CH$_{3}$CN (Table \ref{tab:mm_line_obs}) are indicated by blue vertical bars.}
\label{fig:ch3cn_fit_example}
\end{figure}

	The temperature of the cold component is estimated using CH$_{3}$CN line emission covered by the CORE+ project. In each spatial pixel, the CH$_{3}$CN $J = 5-4$ $K$-ladder is modeled using \texttt{XCLASS} \citep{XCLASS}. An example toward the location of the mmA1 source is shown in Fig. \ref{fig:ch3cn_fit_example}. In these examples, we trace for source mmA1 temperature components at 50\,K, 400\,K, and 1\,400\,K using the H$_{2}$ and CH$_{3}$CN lines as thermometers.

\setlength{\tabcolsep}{4pt}
\begin{table}[!htb]
\caption{Line integrated intensities of H$_{2}$ (extinction-corrected, adopting values of $A_K=7$, 5, and 3 mag) derived from a Gaussian fit to the observed line profiles (Sect. \ref{sec:H2_modeling}).}
\label{tab:H2_line_flux}
\centering
\begin{tabular}{lrrrrrrr}
\hline\hline
Position & \multicolumn{7}{c}{Line integrated intensity of H$_{2}$ 0-0} \\ \cline{2-8}
 & S(1) & S(2) & S(3) & S(4) & S(5) & S(6) & S(7)\\
 & \multicolumn{7}{c}{($10^{-15}$\,W\,m$^{-2}$\,arcsec$^{-2}$)}\\
\hline
\multicolumn{8}{c}{$A_K=7$ mag}\\
mmA1 & 12.0 & 4.8 & 12.4 & 3.3 & 3.1 & 0.9 & 2.2 \\ 
mmB & 11.7 & 5.5 & 10.7 & 4.2 & 5.1 & 1.5 & 3.3 \\ 
A/mmA2 & 18.1 & 7.2 & 19.2 & 5.5 & 5.7 & 1.5 & 3.6 \\ 
B & 21.7 & 8.9 & 18.3 & 7.2 & 7.1 & 1.9 & 4.5 \\ 
S1 & 22.6 & 10.1 & 30.7 & 9.9 & 14.8 & 5.1 & 10.0 \\ 
S2 & 8.6 & 3.4 & 3.9 & 1.6 & 1.5 & 0.5 & 1.0 \\ 
S3 & 4.8 & 2.0 & 1.7 & 2.1 & 1.0 & 0.4 & 1.4 \\ 
S4 & 10.1 & 4.2 & 7.5 & 2.0 & 2.3 & 0.9 & 1.8 \\ 
\hline
\multicolumn{8}{c}{$A_K=5$ mag}\\
mmA1 & 4.7 & 1.9 & 3.3 & 1.4 & 1.4 & 0.4 & 0.9 \\ 
mmB & 4.6 & 2.2 & 2.8 & 1.7 & 2.2 & 0.6 & 1.4 \\ 
A/mmA2 & 7.1 & 2.9 & 5.1 & 2.3 & 2.5 & 0.6 & 1.5 \\ 
B & 8.5 & 3.6 & 4.8 & 3.0 & 3.1 & 0.8 & 1.9 \\ 
S1 & 8.9 & 4.0 & 8.1 & 4.2 & 6.5 & 2.1 & 4.2 \\ 
S2 & 3.4 & 1.4 & 1.0 & 0.7 & 0.7 & 0.2 & 0.4 \\ 
S3 & 1.9 & 0.8 & 0.4 & 0.9 & 0.5 & 0.2 & 0.6 \\ 
S4 & 4.0 & 1.7 & 2.0 & 0.9 & 1.0 & 0.4 & 0.8 \\ 
\hline
\multicolumn{8}{c}{$A_K=3$ mag}\\
mmA1 & 1.8 & 0.8 & 0.9 & 0.6 & 0.6 & 0.1 & 0.4 \\ 
mmB & 1.8 & 0.9 & 0.8 & 0.7 & 1.0 & 0.2 & 0.6 \\ 
A/mmA2 & 2.8 & 1.2 & 1.3 & 1.0 & 1.1 & 0.3 & 0.6 \\ 
B & 3.3 & 1.4 & 1.3 & 1.3 & 1.4 & 0.3 & 0.8 \\ 
S1 & 3.5 & 1.6 & 2.1 & 1.7 & 2.9 & 0.9 & 1.8 \\ 
S2 & 1.3 & 0.5 & 0.3 & 0.3 & 0.3 & 0.1 & 0.2 \\ 
S3 & 0.7 & 0.3 & 0.1 & 0.4 & 0.2 & 0.1 & 0.2 \\ 
S4 & 1.6 & 0.7 & 0.5 & 0.4 & 0.4 & 0.1 & 0.3 \\ 
\hline 
\end{tabular}
\tablefoot{Transition properties are listed in Table \ref{tab:MIRI_line_obs}.}
\end{table}

\setlength{\tabcolsep}{5pt}
\begin{table*}[!htb]
\caption{Fit results from the H$_{2}$ excitation diagram analysis with \texttt{pdrtpy} (Sect. \ref{sec:H2_modeling}) with a warm and hot component, adopting values of $A_K=7$, 5, and 3 mag.}
\label{tab:H2_fit_results}
\centering
\renewcommand{\arraystretch}{1.2}
\begin{tabular}{lrrrr}
\hline\hline
Position & \multicolumn{2}{c}{\underline{Warm component}} & \multicolumn{2}{c}{\underline{Hot component}}\\
 & Temperature & Column density & Temperature & Column density \\
 & $T_\mathrm{warm}$ & log $N_\mathrm{warm}$ & $T_\mathrm{hot}$ & log $N_\mathrm{hot}$ \\
 & (K) & (cm$^{-2}$) & (K) & (cm$^{-2}$)\\ 
\hline
\multicolumn{5}{c}{$A_K=7$ mag}\\
mmA1 & $414\pm44$ & $23.70\pm0.22$ & $1407\pm522$ & $21.21\pm0.74$ \\ 
mmB & $346\pm56$ & $23.89\pm0.30$ & $1033\pm169$ & $22.00\pm0.41$ \\ 
A/mmA2 & $409\pm51$ & $23.88\pm0.24$ & $1240\pm385$ & $21.65\pm0.69$ \\ 
B & $357\pm67$ & $24.10\pm0.35$ & $1015\pm220$ & $22.17\pm0.56$ \\ 
S1 & $371\pm54$ & $24.09\pm0.25$ & $1100\pm164$ & $22.37\pm0.35$ \\ 
S2 & $294\pm53$ & $23.94\pm0.41$ & $970\pm177$ & $21.65\pm0.47$ \\ 
S3 & $258\pm108$ & $23.89\pm0.97$ & $1011\pm306$ & $21.58\pm0.71$ \\ 
S4 & $358\pm39$ & $23.77\pm0.23$ & $1214\pm259$ & $21.41\pm0.46$ \\ 
\hline
\multicolumn{5}{c}{$A_K=5$ mag}\\
mmA1 & $329\pm65$ & $23.53\pm0.39$ & $991\pm195$ & $21.53\pm0.51$ \\ 
mmB & $285\pm61$ & $23.73\pm0.45$ & $987\pm145$ & $21.73\pm0.36$ \\ 
A/mmA2 & $305\pm64$ & $23.80\pm0.42$ & $937\pm145$ & $21.89\pm0.41$ \\ 
B & $276\pm69$ & $24.02\pm0.53$ & $930\pm145$ & $22.01\pm0.40$ \\ 
S1 & $298\pm50$ & $23.94\pm0.33$ & $1035\pm110$ & $22.12\pm0.25$ \\ 
S2 & $255\pm57$ & $23.76\pm0.55$ & $955\pm171$ & $21.31\pm0.45$ \\ 
S3 & $238\pm107$ & $23.62\pm1.11$ & $1057\pm344$ & $21.11\pm0.72$ \\ 
S4 & $295\pm47$ & $23.61\pm0.35$ & $1044\pm167$ & $21.34\pm0.38$ \\ 
\hline
\multicolumn{5}{c}{$A_K=3$ mag}\\ 
mmA1 & $270\pm68$ & $23.40\pm0.55$ & $949\pm161$ & $21.26\pm0.43$ \\ 
mmB & $259\pm72$ & $23.47\pm0.62$ & $1022\pm184$ & $21.28\pm0.42$ \\ 
A/mmA2 & $264\pm67$ & $23.61\pm0.56$ & $941\pm145$ & $21.51\pm0.39$ \\ 
B & $252\pm75$ & $23.77\pm0.69$ & $962\pm173$ & $21.56\pm0.44$ \\ 
S1 & $268\pm59$ & $23.69\pm0.47$ & $1071\pm143$ & $21.68\pm0.29$ \\ 
S2 & $235\pm64$ & $23.50\pm0.70$ & $989\pm208$ & $20.86\pm0.50$ \\ 
S3 & $226\pm109$ & $23.32\pm1.26$ & $1122\pm418$ & $20.62\pm0.77$ \\ 
S4 & $260\pm53$ & $23.40\pm0.48$ & $1041\pm179$ & $20.97\pm0.40$ \\ 
\hline 
\end{tabular}
\end{table*}

\end{appendix}

\end{document}